\documentclass[twocolumn,twoside,slac_two]{revtex4}
\usepackage{graphicx}
\usepackage{fancyhdr}
\input babarsym.tex
\def\CP{$ C \! P$ } 
\def\CPn{$ C \! P$}

\def\ra{\rightarrow}

\def\babar{\mbox{\slshape B\kern-0.1em{\smaller A}\kern-0.1em B\kern-0.1em{\smaller A\kern-0.2em R}}~} 
\def\babarn{\mbox{\slshape B\kern-0.1em{\smaller A}\kern-0.1em B\kern-0.1em{\smaller A\kern-0.2em R}}}

\begin{document}

\title{Exclusive ${\bf b \rightarrow s (d) \ell \ell}$ Decays}

%

\author{G. Eigen  (representing the \babar collaboration)}
\affiliation{University of Bergen, Bergen, Norway }

\begin{abstract}
New \babar measurements are presented for the exclusive rare decays $B \ra K^{(*)} \ell^+ \ell^-$ including branching fractions, isospin asymmetries, direct \CP violation, and lepton flavor universality for dilepton masses below and above the $J/\psi$ resonance. Unexpectedly large isospin asymmetries are observed in both $K \ell^+ \ell^-$ and $K^* \ell^+ \ell^-$ decays. For the combined $K \ell^+ \ell^-$ and $K^* \ell^+ \ell^-$ data a $3.9\sigma$ significant deviation from the SM prediction is found. Furthermore, recent \babar results from an angular analysis in $B \ra K^* \ell^+ \ell^-$ are reported in which both the $K^*$ longitudinal polarization and the lepton forward-backward asymmetry are measured for dilepton masses below and above the $J/\psi$ resonance. Finally, results of recent searches for $B \ra \pi \ell^+ \ell^-$ from Belle and $B \ra K^{(*)} \nu \bar \nu$ from \babar are summarized. 

\end{abstract}

\maketitle

\thispagestyle{fancy}


\section{Introduction}
\begin{figure}[h]
\centering
\includegraphics[width=85mm]{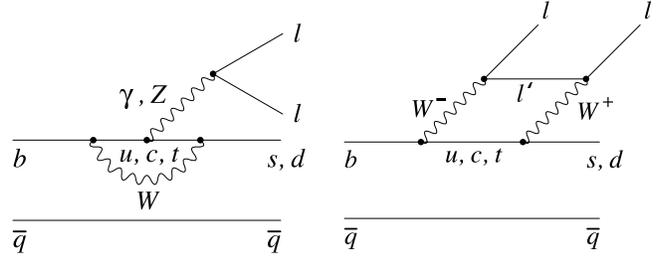}
\caption{Lowest-order SM processes for $b \ra s (d) \ell \ell$; (left) electroweak penguin loops and (right) weak box diagram. If $\ell $ is a charged lepton (neutrino), $\ell '$ is a neutrino (charged lepton).}
 \label{fig:bsll}
\end{figure}

The flavor-changing neutral-current (FCNC) processes, $b \ra s (d) \ell \ell$,  provide an interesting hunting ground to look for new-physics phenomena. In the Standard Model (SM), FCNC transitions are forbidden at tree level, but they are allowed to proceed via electroweak-loop and weak-box diagrams. Lowest-order processes are depicted in Figure~\ref{fig:bsll}. The electroweak loops consist of a W-boson and a $t,~c$ or $u$-quark, where the $t$-quark contribution dominates. One of the loop particles radiates a photon or Z-boson to conserve momentum. In the weak-box diagram the $b$ to $s(d)$ transition occurs via the emission of two $W$ bosons. The emitted lepton pair is $\epem$, $\mumu$, $\tau^+ \tau^-$, or $\nu \bar \nu$. Final states with a $\tau^+ \tau^-$ pair, however, are difficult to measure because of the missing neutrinos and are not discussed further.

In chapter 2, we discuss properties of  the exclusive decays $B \ra K^{(*)} \ell^+ \ell^-$ where $\ell^+ \ell^-$ is either $\epem$ or $\mumu$ and introduce the observables we measure. In chapter 3, we focus on \babar measurements of  $B \ra K^{(*)} \ell^+ \ell^-$ branching fractions, isospin asymmetries, direct \CP violation and lepton flavor universality after discussing the analysis strategy. In chapter 4, we discuss the \babar angular analysis of $B \ra K^* \ell^+ \ell^-$ decays. In chapter 5, we present the latest branching fraction upper limits of $B \ra \pi \ell^+ \ell^-$ decays. In chapter 6, we show results of recent searches for $B \ra K \nu \bar \nu$ and   $B \ra K^* \nu \bar \nu$ before concluding in chapter 7.

\section{Properties of $B \rightarrow K^{(*)} \ell^+ \ell^- $ Decays}
\label{theory}

The SM calculations are based on a low-energy effective Hamiltonian \cite{buchalla} that factorizes perturbatively-calculable short-distance contributions expressed by Wilson coefficients, $C_i$, from the non-perturbative long-distance contributions of the $b \rightarrow s$ transition operators, ${\cal O}_i$. 
In lowest order, dominant contributions result from the magnetic dipole operator ${\cal O}_7$, the vector-current operator  ${\cal O}_9$ and the axial-vector-current operator ${\cal O}_{10}$. While ${\cal O}_7$ represents the photon penguin diagram, ${\cal O}_9$ and ${\cal O}_{10}$ result from linear combinations of the weak penguin and weak box diagrams. Thus, the relevant Wilson coefficients for these modes are $C_7$, $C_9$ and $C_{10}$. QCD effects, however, introduce operator mixing. Besides $\alpha_s$ corrections, the Wilson coefficients also receive contributions from other operators  \cite{misiak, greub, hiller03}. It is customary to absorb these by defining effective Wilson coefficients $\widetilde C_i$. The Wilson coefficients are functions of the renormalization scale $\mu$ and the squared dilepton mass, $q^2 $. For low values of $q^2$, $\rm \mu= 4.6~GeV$ and a top quark mass $\widehat m_t(\widehat m_t)=167~\rm GeV/c^2$ in the $\overline{MS}$ renormalization scheme, the SM predictions at next-to-next-to-leading order (NNLO) yield $\widetilde C_7 = -0.31 $, $\widetilde C_9 =4.21 $ and $ \widetilde C_{10} = -4.31$ \cite{beneke01, misiak, greub}. The magnitude of $\widetilde C_7$ is constrained experimentally by ${\cal B}(B \ra X_s \gamma)$ \cite{misiak93}.

\begin{figure}[h]
\centering
\includegraphics[width=38mm]{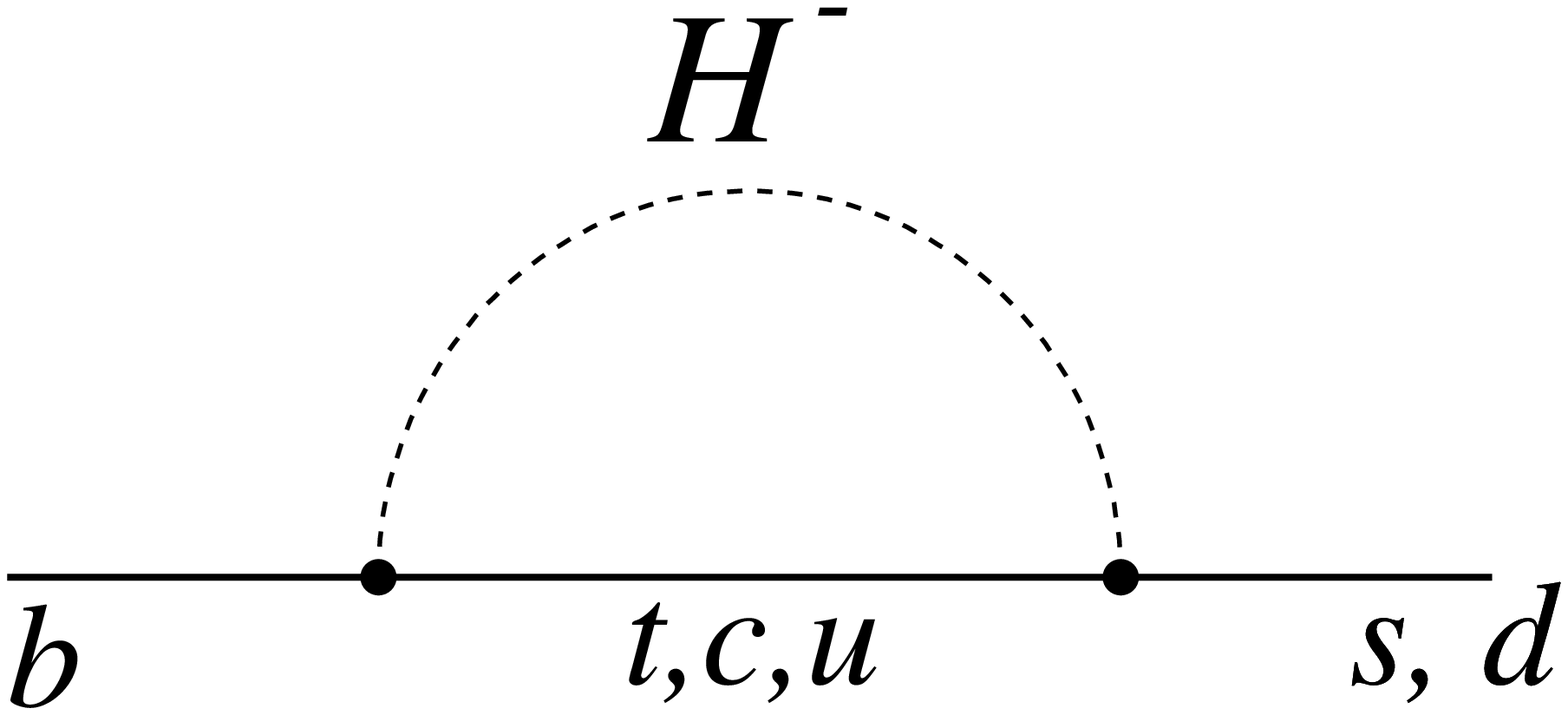}
\includegraphics[width=38mm]{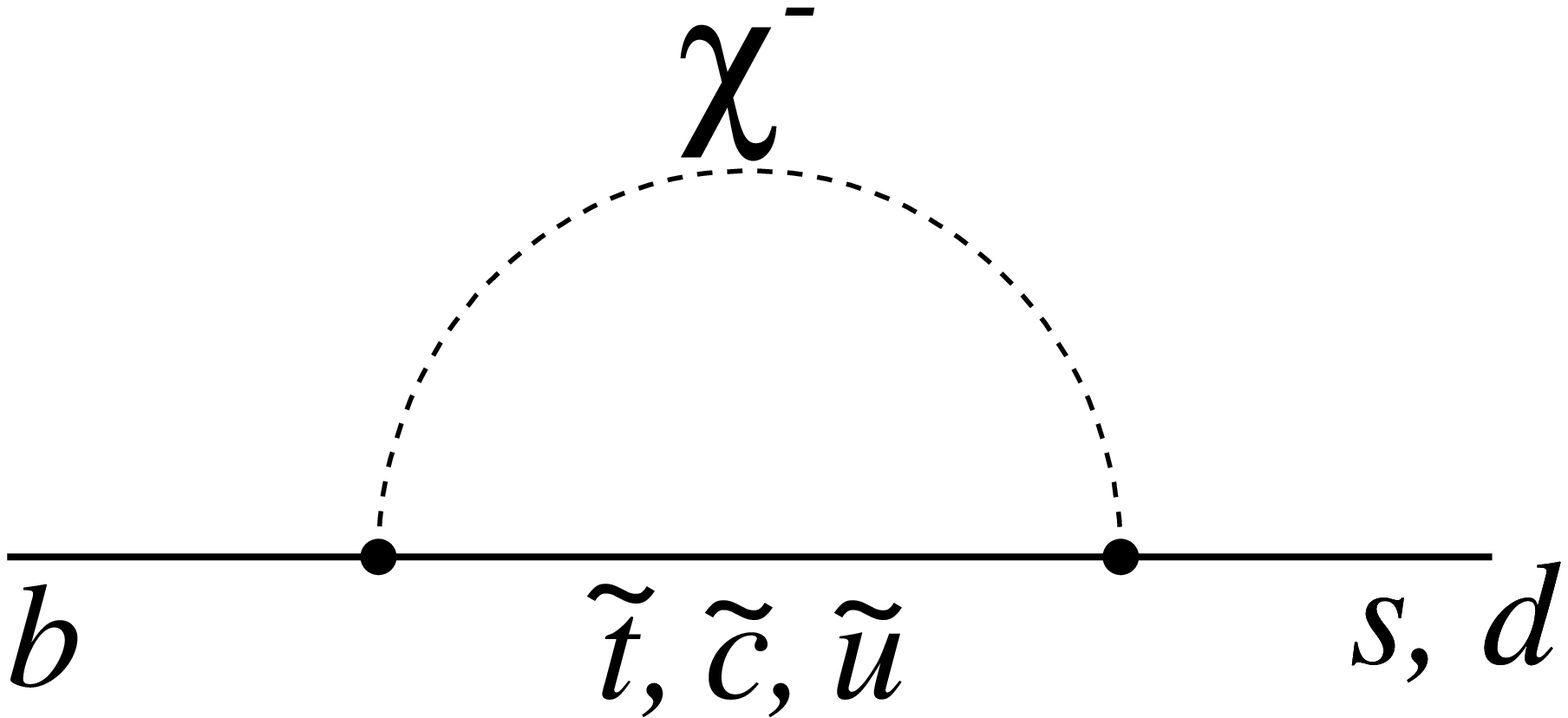}
\includegraphics[width=38mm]{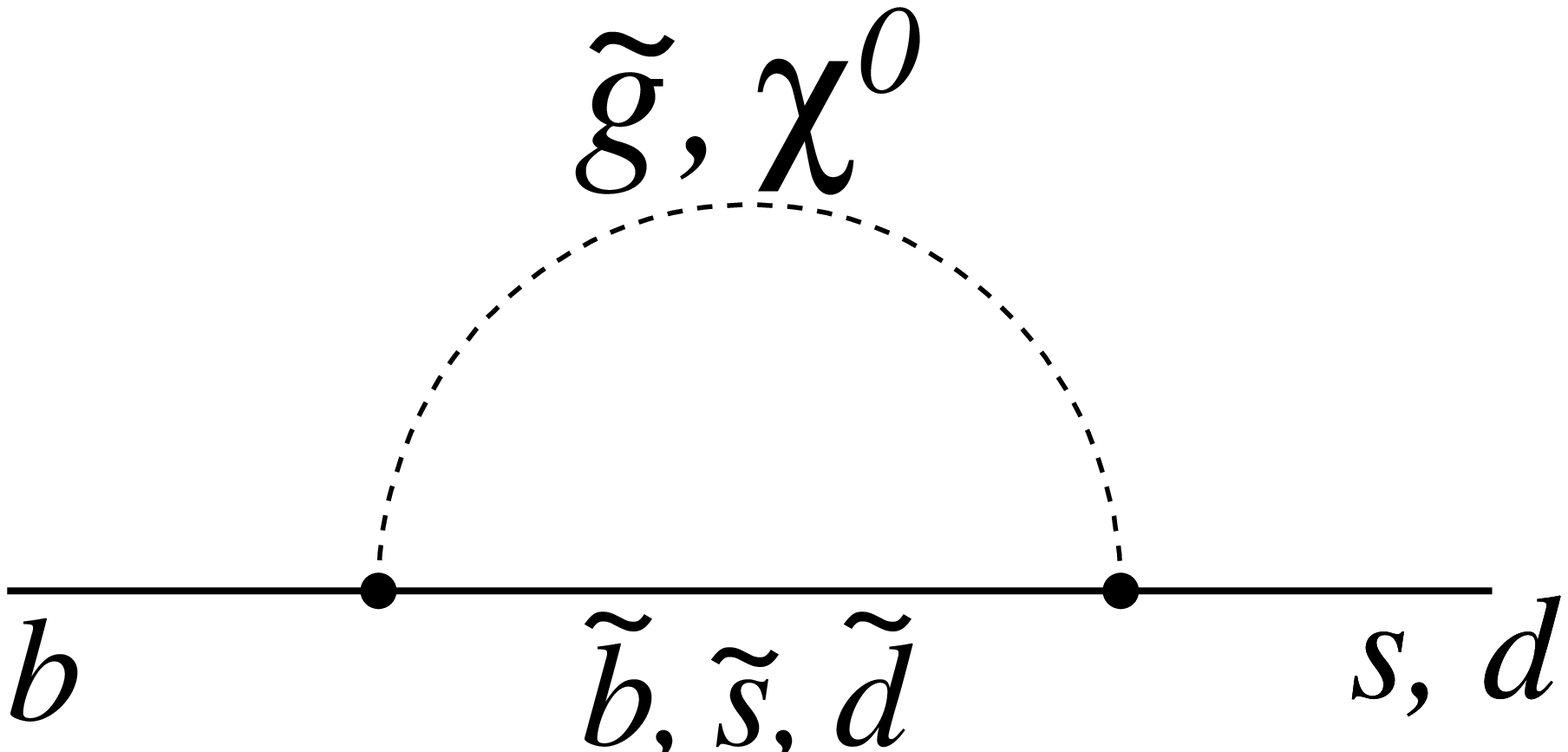}
\caption{Examples of new-physics contributions in loop processes, (top left) a charged Higgs boson with $u,c,t$ quarks, (top right) a chargino with $u,c,t$ squarks, and (bottom) a neutralino or gaugino with $d,s,b$ squarks.}
 \label{fig:np}
\end{figure}

Since FCNC transitions are suppressed in the SM, loops and box diagrams from processes beyond the SM may yield non-negligible contributions. For example, new contributions may arise from loops containing a charged Higgs boson or supersymmetric (SUSY) particles \cite{NewPhysics, feldmann02, yan}. Examples are shown in Figure~\ref{fig:np}. New physics contributions will modify the effective Wilson coefficients from their SM expectations. In addition, scalar and pseudoscalar couplings that are absent in the SM may modify the decay rate \cite{hiller03}. In order to have the ability for uncovering new physics phenomena, the SM predictions need to have sufficient precision. Most recent calculations use QCD factorization to separate the short-distance physics from the long-distance effects \cite{beneke01,beneke05}. The effective Wilson coefficients entering the short-distance part are calculated in the NNLO approximation. The long-distance effects including the hadronization process are expressed in terms of hadronic matrix elements of the $b \ra s$ transition operators between the initial $B$ and the $K^{(*)}$ final states. Since the hadronic matrix elements cannot be calculated from first principles, they are parameterized in terms of form factors \cite{ali} that are calculated with the help of light-cone sum rules (LCSR) \cite{ball} or in soft-collinear effective theory (SCET) \cite{defazio}. Though the form-factor calculations include next-to-leading order (NLO) QCD corrections they bear large theoretical uncertainties that are presently the dominant uncertainties in the SM predictions for exclusive decays. Thus, it is important to measure many different observables in different inclusive and exclusive electroweak penguin processes in order to extract meaningful results for moduli and arguments of the effective  Wilson coefficients.

\begin{figure}[h]
\centering
\includegraphics[width=80mm]{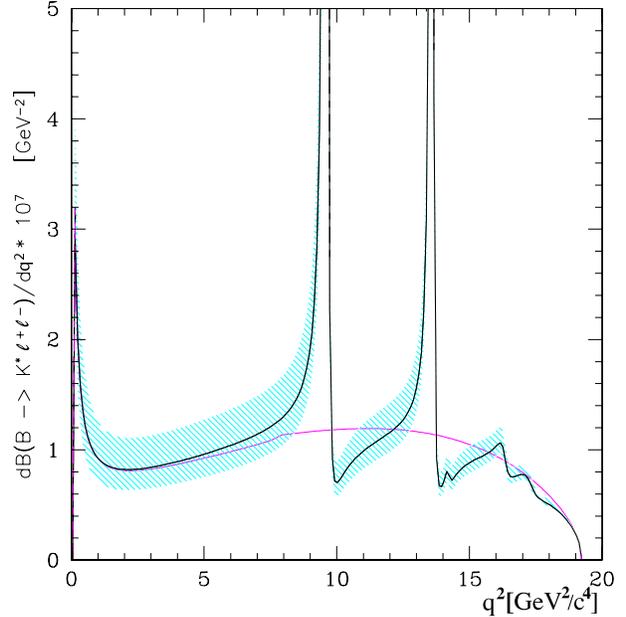}
\caption{Predicted $B \ra K^* \ell^+ \ell^-$ branching fraction in the SM as a function of $q^2$ with (dark curve) and without (light curve) contributions from charmonium resonances \cite{ali}.  The shaded region indicates the SM uncertainties.}
 \label{fig:spectrum}
\end{figure}

Figure~\ref{fig:spectrum} shows the $q^2$ dependence of the $B \ra K^* \ell^+ \ell^-$ branching fraction in the SM \cite{ali}. The overall shape is determined by the $q^2$ dependence of $\widetilde C_9(q^2)$ except for the low $q^2$ region that is dominated by the $1/q^2$ term originating from $B \ra K^* \gamma$. The singularity at $q^2=0$, which is not present in $B \ra K \ell^+ \ell^-$, is cut off by the finite $m_\ell$ masses. Thus, ${\cal B}(B \ra K^* e^+ e^-)$ is expected to be $\sim25\%$ larger in the SM than ${\cal B}(B \ra K^* \mu^+ \mu^-)$, where the increase in branching fraction comes just from the extended $q^2 < 4 m_\mu^2$ region. In addition, the hadronic decays $B \ra J/\psi K^*$ and $B \ra \psi(2S) K^*$ with $J/\psi, \psi(2S) \ra \ell^+ \ell^-$ interfere with signal modes. Since the branching fractions of the charmonium modes are more than two orders of magnitude larger than those of signal modes, we need to exclude sufficiently large $q^2$ regions in the vicinity of the charmonium resonances to remove them (see section \ref{kll}).

Besides partial and total branching fractions, asymmetries are of great interest in testing the SM, in particular for exclusive decays since many uncertainties in both predictions and measurements cancel \cite{kruger}. For $B \ra K \ell^+ \ell^-$ and $B \ra K^* \ell^+ \ell^-$ we have performed new measurements of the isospin asymmetries, direct \CP violation and lepton flavor universality.

\begin{figure}[h]
\centering
\includegraphics[width=80mm]{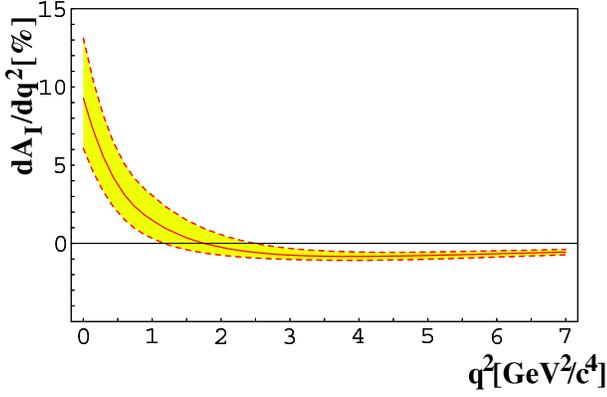}
\caption{Predicted $q^2$ dependence of isospin asymmetry ${\cal A}_I(q^2)$ in the SM in the low $q^2$ region \cite{feldmann02}.}
 \label{fig:iso-sm}
\end{figure}

The \CPn-averaged isospin asymmetry is defined as

\begin{eqnarray}
{\cal A}^{K^{(*)}}_{I} \equiv
\frac
{{\cal B}(\Bz \to K^{(*)0}\ellell) - r{\cal B}(\B^{\pm} \to K^{(*)\pm}\ellell)}
{{\cal B}(\Bz \to K^{(*)0}\ellell) + r{\cal B}(\B^{\pm} \to K^{(*)\pm}\ellell)}
\nonumber \\
\;
\end{eqnarray}
\noindent 
where $r= \tau_0/\tau_+=1/(1.071 \pm 0.009)$ \cite{hfag} is the ratio of $B^0$ and $B^+$ lifetimes. Figure~\ref{fig:iso-sm} shows the $q^2$ dependence of ${\cal A}_I^{K^*}$ in the SM. For $q^2 \ra 0$ the isospin asymmetry is  $6-13\%$. With increasing $q^2$, ${\cal A}_I$ decreases approaching a value $\sim -1\%$  for $q^2>2.5~\rm GeV^2/c^4$ \cite{feldmann02}.

The direct \CP asymmetries 
\begin{eqnarray}
{\cal A}_{CP}^{K^{(*)}} \equiv
\frac
{{\cal B}(\overline{B} \rightarrow \overline{K}^{(*)}\ellell) - {\cal B}(B \rightarrow K^{(*)}\ellell)}
{{\cal B}(\overline{B} \rightarrow \overline{K}^{(*)}\ellell) + {\cal B}(B \rightarrow K^{(*)}\ellell)}
\end{eqnarray}
\noindent may provide useful constraints on non-SM physics~\cite{kruger2}. The SM ${\cal A}_{CP}$ predictions for modes studied here are very small, ${\cal O}(10^{-3})$, and new physics at the electroweak scale may provide significant enhancements \cite{bobeth08}.

The ratios of rates to $\mu^+ \mu^-$ and $e^+ e^-$ final states
\begin{eqnarray}
{\cal R}_{K^{(*)}} \equiv
\frac
{{\cal B}(B \ra K^{(*)} \mu^+ \mu^-)}
{{\cal B}(B \ra K^{(*)} e^+ e^-)}
\end{eqnarray}
\noindent are sensitive to the presence of a neutral SUSY Higgs boson~\cite{yan}. In the SM, ${\cal R}_{K}$ is expected to be unity modulo a small correction accounting for differences in phase space ~\cite{hiller03}. ${\cal R}_{K^*}$ should also be close to unity for $m_{\ell\ell} \geq 2m_{\mu}$. Due to the $1/q^2$ dependence of the photon penguin contribution, however, there is a significant rate enhancement in $K^* e^+ e^-$ modes for $m_{\epem}<2m_{\mu}$. The expected SM value of ${\cal R}_{K^*}$ including this region is 0.75. In order to test this predicted rate enhancement, we fit the $K^*$
dataset in the entire $q^2$ and in the low $q^2$ region with and without inclusion of events in the $q^2 <0.1~ \rm GeV^2/c^4$ region. Theoretical uncertainties for ${\cal R}_{K^{(*)}}$ predictions in the SM are just a few percent. For example, in two-Higgs-doublet models the presence of a SUSY Higgs might give $\sim 10\%$ corrections to ${\cal R}_{K^{(*)}}$ for large $\tan \beta$ \cite{yan}.

Besides $q^2$, the $B \ra K^* \ell^+ \ell^-$ differential decay rate also depends on three angles,
$\theta_K, \theta_\ell$, and $\chi$, where $\theta_K$ is the angle between the $K$ and the $B$ momenta in the $K^*$  rest frame, $\theta_\ell$ is the angle between the $\ell^+ (\ell^-)$ and the $B(\bar B)$ momentum in the $\ell^+ \ell^-$ rest frame, and $\chi$ is the angle between the $K^*$ and $\ell^+\ell^-$ decay planes.
The present data sample is not large enough to perform a full three-dimensional angular analysis. Thus, we fit the one-dimensional angular distributions \cite{kruger2}

\begin{eqnarray} 
W(\cos \theta_k) & = &\frac{3}{2} {\cal F}_L \cos^2 \theta_K
+ \frac{3}{4} (1- {\cal F}_L) \sin^2 \theta_K,  \nonumber \\
W(\cos \theta_\ell) & = &\frac{3}{4} {\cal F}_L \sin^2 \theta_\ell
+ \frac{3}{8} (1 -{\cal F}_L)(1 + \cos^2 \theta_\ell)  \nonumber \\
&  +& {\cal A}_{FB}\cos \theta_\ell .
\label{eq-theta}
\end{eqnarray}

\begin{figure}[h]
\centering
\includegraphics[width=75mm]{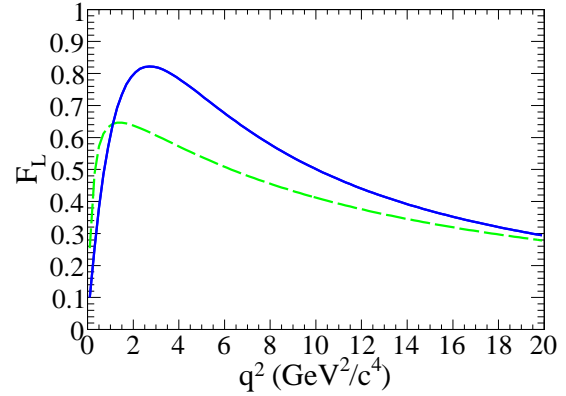}
\caption{Predicted $q^2$ dependence of the $K^*$ longitudinal polarization in the SM (solid) and the flipped-sign $\widetilde C_7$ model (dashed).}
 \label{fig:fl}
\end{figure}

The parameter ${\cal F}_L$ is the $K^*$ longitudinal polarization. The second parameter, ${\cal A}_{FB}$, is the lepton forward-backward asymmetry. Both parameters are functions of $q^2$. Figures~\ref{fig:fl}  shows the $q^2$ dependence of the $K^*$ longitudinal polarization. In addition to the SM prediction, the distribution for the flipped-sign $\widetilde C_7$ model is shown for which the sign of  $\widetilde C_7$ is opposite to that in the SM.


The $q^2$ dependence of lepton forward-backward asymmetry is given by

\begin{eqnarray}
  \frac{{\rm d} {\cal A}_{FB}}{{\rm d} q^2}& \propto 
  -{\cal R}e [( \widetilde C_{9}(q^2)) \widetilde C_{10}(q^2) V(q^2) A_1(q^2) \nonumber \\
&  +  \frac{M_B m_b}{q^2} 
 \widetilde C_{7}(q^2) \widetilde C_{10}(q^2) (V(q^2) 
 T_2(q^2)  \nonumber \\
&  (1-\frac{m_{K^*}}{M_B}) 
 + A_1(q^2) T_1(q^2) (1+\frac{m_{K^*}}{M_B}))].  \;
\end{eqnarray}

\begin{figure}[h]
\centering
\includegraphics[width=80mm]{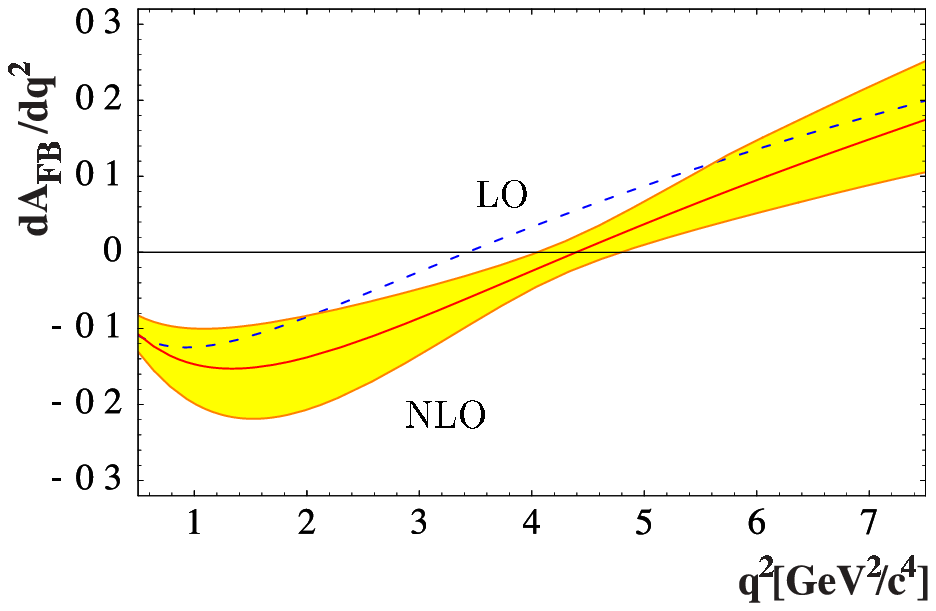}
\caption{The low $q^2$ region of ${\cal A}_{FB}$ for $B \ra K^* \ell^+ \ell^-$ \cite{feldmann02}.}
 \label{fig:afb-0k}
\end{figure}

\begin{figure}[h]
\centering
\includegraphics[width=80mm]{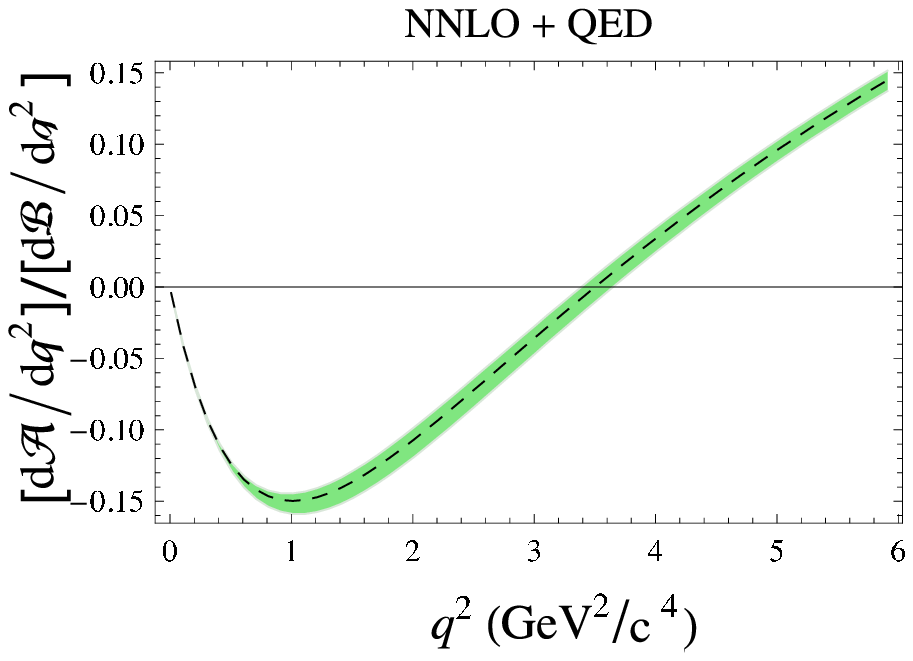}
\caption{The low $q^2$ region of ${\cal A}_{FB}$ for $B \ra X_s \ell^+ \ell^-$ \cite{huber}.}
 \label{fig:afb-0xs}
\end{figure}

The functions $A_1(q^2), V(q^2), T_1(q^2)$ and $T_2(q^2)$ are $q^2$-dependent form factors. The shape of the lepton forward-backward asymmetry results from different $q^2$ dependences of the interference terms $-\widetilde C_9(q^2) \widetilde C_{10}(q^2)$ and $-\widetilde C_7(q^2) \widetilde C_{10}(q^2)/q^2$. In the SM, the first (second) term is positive (negative) and increases (decreases) with $q^2$. Thus, ${\cal A}_{FB}$ is negative at low $q^2$, crosses zero and remains positive for large values of $q^2$ \cite{ali, kruger2, stewart}. For $B \ra K^* \ell^+ \ell^-$ the zero crossing is predicted in NLO at $q^2_0= 4.2\pm 0.6$. Figure~\ref{fig:afb-0k} shows the low $q^2$ region for $B \ra K^* \ell^+ \ell^-$ \cite{feldmann02}.  The uncertainty of $q^2_0$ is dominated by uncertainties in the form-factor calculations. This is different for inclusive modes. Recent calculations of ${\cal A}_{FB}$ in $B \ra X_s \ell^+ \ell^-$ at order ${\cal O}(\alpha^2_s)$ (NNLO) that include in addition electromagnetic corrections predict a $q^2$ dependence shown in Figure~\ref{fig:afb-0xs} \cite{huber}. The zero crossing is expected at $q^2_0 =3.5\pm 0.12$.

\begin{figure}[h]
\centering
\includegraphics[width=80mm]{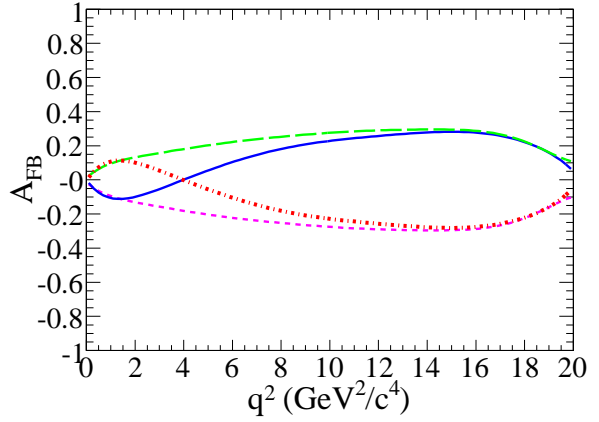}
\caption{Predicted $q^2$ dependence of the lepton forward-backward asymmetry in the SM (solid), the flipped-sign $\widetilde C_7$ model (dashed), the flipped-sign $\widetilde C_9 \widetilde C_{10}$ model (dotted) and the mirror image of the SM (dash-dotted) \cite{ali}.}
 \label{fig:afb}
\end{figure}

Figure~\ref{fig:afb} shows the SM prediction of ${\cal A}_{FB}(q^2)$ in the entire $q^2$ region. New physics may change both magnitude and phase of the Wilson coefficients \cite{kim2lu} yielding shapes that differ from those in the SM. We consider three simple examples, also shown in Figure~\ref{fig:afb}, in which the signs of the two interference terms are reversed ($i.e.$ a phase change by $\pi$) \cite{ali, kruger, kruger2, hou, buchalla2, bobeth01, ali2}. Changing the sign in $\widetilde C_7$, yields a positive ${\cal A}_{FB}$ for all values of $q^2$, whereas a reversed sign in $\widetilde C_9(q^2) \widetilde C_{10}(q^2)$ yields a negative ${\cal A}_{FB}$ in the entire $q^2$ region. If the sign of both interference terms is flipped, the resulting distribution has the negative mirror image of the SM distribution.

\section{Study of  $B \rightarrow K^{(*)} \ell^+ \ell^-$}
\label{kll}

We have studied the exclusive decays  $B \rightarrow K \ell^+ \ell^- $ and  $B \rightarrow K^* \ell^+ \ell^- $ using $384$ million $\BB$ pairs collected at the $\FourS$ resonance with the \babar detector~\cite{BaBarDetector} at the \pep2\ asymmetric-energy $\epem$ collider. We reconstruct ten individual final states, in which a $\rm K^+$, $\rm K^0_S (\ra \pi^+ \pi^-)$, $\rm K^{*0}(892) (\ra K^+ \pi^-)$ or a $\rm K^{*+}(892) (\ra K^+ \pi^0 ~or~K^0_s \pi^+) $ recoils against an $\epem$ or a $\mumu$ pair. We require good particle identification (PID) for $e^\pm, \mu^\pm, K^\pm$ and $\pi^\pm$. We select electrons (muons) with momenta $p > 0.3~(0.7) \gevc$ in the laboratory frame. We merge photons consistent with $ e^\pm$ decay radiation or bremsstrahlung with the corresponding $e^\pm$ and veto events in which the $\epem$ pair is consistent with a photon conversion. We select  $K^0_S$ candidates from $\pi^+ \pi^-$ final states for which the invariant mass is consistent with the nominal $K^0$ mass and the flight distance from the interaction point is larger than three times its uncertainty. We form $\pi^0$ candidates from two photons with energies larger than $50 \mev$ having an invariant mass  of $115 \mevcc < m_{\gamma \gamma}  < 155 \mevcc$. We require $\Kstar(892)$ candidates to have an invariant mass $0.82 < m_{K\pi} < 0.97 \gevcc$.
Note that we imply charge conjugation throughout this article unless otherwise noted.

We split the data set into two $q^2$ regions, low $q^2$ ($0.1 <  q^2 < 7.02 ~\rm GeV^2/c^4$) and high $q^2$ ($\rm 10.24   < q^2< 12.96~GeV^2/c^4 $ and $q^2 > 14.06 ~GeV^2/c^4$). For the $K^*\ell^+ \ell^-$ mode, we also report results that include the region of $\rm q^2 < 0.1~GeV^2/c^4$. We use two kinematic variables to select signal events, $\mes=\sqrt{s/4 -p^{*2}_B}$ and $\Delta E = E_B^* - \sqrt{s}/2$, where $p^*_B$ and $E_B^*$ are the $B$ momentum and energy in the $\Upsilon(4S)$ center-of-mass (CM) frame, and $\sqrt{s}$ is the total CM energy. We extract signal yields from a one-dimensional fit to the $\mes$ distribution for $\mes > 5.2 \gevcc$, after a selection on $\Delta E$: $-0.07<\Delta E<0.04$ ($-0.04<\Delta E<0.04$) $\gev$ for $e^+e^-$ ($\mu^+\mu^-$) events in the low $q^2$ region, and $-0.08<\Delta E<0.05$ ($-0.05<\Delta E<0.05$) $\gev$ for $e^+e^-$ ($\mu^+\mu^-$) events in the high $q^2$ region.

The main backgrounds arise from combinations of leptons from two semileptonic decays ($B \ra X_1 \ell^+ \nu_\ell$ and $\bar B \ra X_2 \ell^- \bar \nu_\ell$ or $B \ra X \ell^+ \nu_\ell$ and $X \ra X^\prime \ell^- \bar \nu_\ell$).  We suppress these  combinatorial backgrounds by using neural networks (NN). For each final state four separate NN are optimized to suppress either continuum or $B\Bbar$ backgrounds in each of the two $q^2$ regions. Inputs to these NN include event shape variables, vertexing information and missing energy, where for each of the ten final states the NN selections are optimized to yield the highest statistical signal significance in the \mes\ signal region ($\mes>5.27 \gevcc$). A potential background contribution arises from $B \to D(K^{(*)} \pi) \pi$ decays, where both pions are misidentified as muons. Therefore, we require the $K^{(*)} \pi$  invariant mass to lie outside the $1.84-1.90 \gevcc$ region to veto this background. The final signal selection efficiencies vary from 3.5\% for $K^+\pi^0\mu^+\mu^-$at low values of $q^2$ to 22\% for $K^+\pi^-e^+e^-$ in the high $q^2$ region.

\begin{figure}[h]
\centering
\includegraphics[width=80mm]{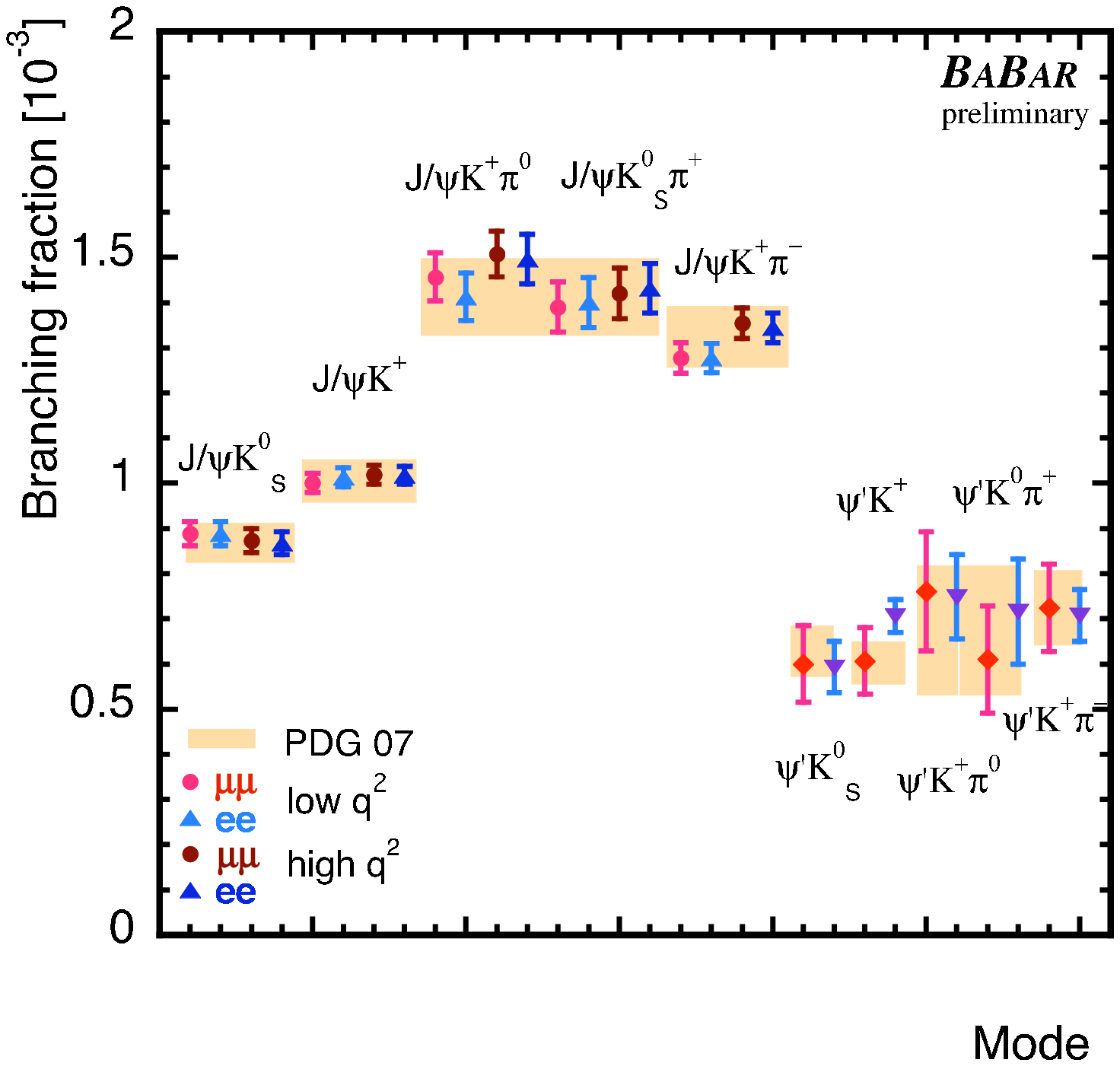}
\caption{Measurement of $J/\psi$ and $\psi(2S)$ branching fractions for the ten final states in comparison to world averages \cite{PDG}. For the $J/\psi~ [\psi(2S)]$ the test is performed in the low $q^2$ and high $q^2$ (entire) regions for $\mumu$ (solid points) and $\epem$ (triangles) final states.}
 \label{fig:charmonium}
\end{figure}

We use the vetoed $J/\psi$ ($7.02 < q^2 < 10.24~\rm GeV^2/c^4$) and $\psi(2S)$ events ($12.96 < q^2 < 14.06~\rm GeV^2/c^4$) as control samples to validate our fit methodology, determine the probability density functions (pdf) for the $\mes$ signal shapes and validate efficiencies.  Figure~\ref{fig:charmonium} shows the $J/\psi$ and $\psi(2S)$ branching fractions we measure for the ten final states in comparison to the world averages~\cite{PDG}. Our measurements agree well with the world averages. For all $J/\psi$ branching fractions and $\psi(2S)$ branching fractions measured in the $\epem$ mode, the uncertainties are smaller than those of the world averages.

We consider systematic uncertainties associated with reconstruction efficiencies, hadronic background parameterization in  $\mu^+ \mu^-$ final states, peaking background contributions obtained from simulations, as well as possible isospin, \CPn, and lepton flavor asymmetries in the background pdfs.
We quantify the efficiency systematic using the vetoed $J/\psi K^{(*)}$ data samples. These include charged track, $\pi^0$ and $K^0_S$ reconstructions, PID, NN event selection, and $\DeltaE$ and $K^{*}$ mass selection. The largest individual systematic errors come from hadronic PID, the characterization of the hadronic background and signal $\mes$ pdf shape. In the rate asymmetries most of these errors cancel at least partially. In general, the systematic uncertainties are very small compared to statistical uncertainties.

\subsection{Branching Fraction Measurements}

\begin{figure}[h]
\centering
\includegraphics[width=80mm]{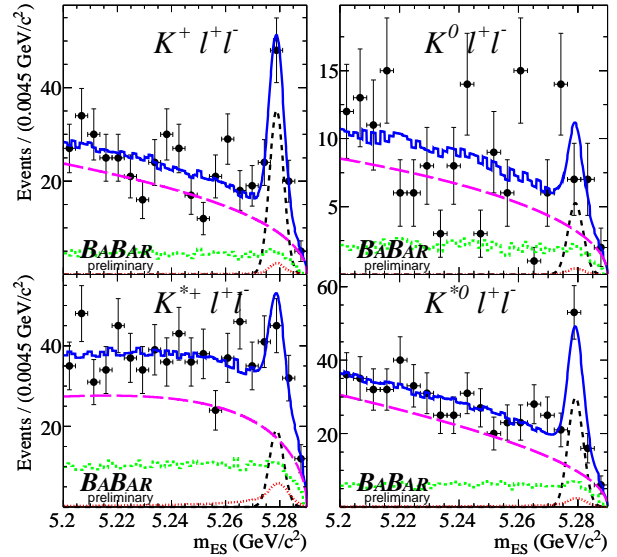}
\caption{Measured $\mes$ distributions in the entire $q^2$ region (points with error bars) for $K^+ \ell^+ \ell^-$  (upper left), $K^0 \ell^+ \ell^-$ (upper right), $K^{*+} \ell^+ \ell^-$ (lower left) and $K^{*0}\ell^+ \ell^-$ (lower right) with fit results superimposed, full fit (blue solid curve), signal contribution (black dashed Gaussian),  combinatorial background (magenta dashed curve), misidentified muons (green dotted histogram), and peaking backgrounds (red dotted Gaussian).}
 \label{fig:mes-all}
\end{figure}

\begin{figure}[h]
\centering
\includegraphics[width=80mm]{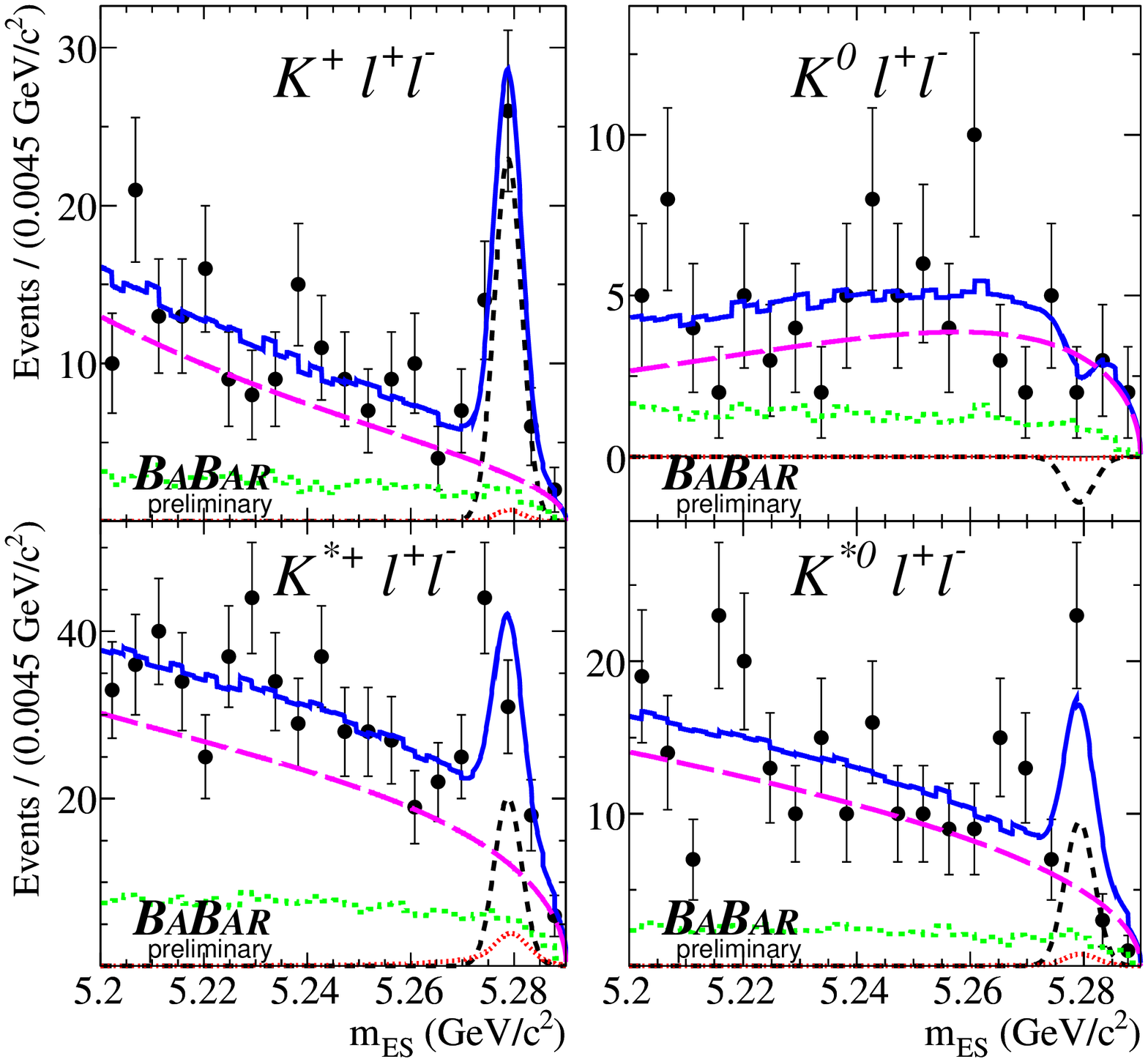}
\caption{Measured $\mes$ distributions in the low $q^2$ region (points with error bars) for $K^+ \ell^+ \ell^-$  (upper left), $K^0\ell^+ \ell^-$ (upper right), $K^{*+}\ell^+ \ell^-$ (lower left) and $K^{*0}\ell^+ \ell^-$ (lower right) with fit results superimposed, full fit (blue solid curve), signal contribution (black dashed Gaussian),  combinatorial background (magenta dashed curve), misidentified muons (green dotted histogram), and peaking backgrounds (red dotted Gaussian). }
 \label{fig:mes-low}
\end{figure}

Figure~\ref{fig:mes-all} shows the $\mes$ distributions for the $K^+, K^0, K^{*+}$ and $K^{*0}$ modes after summing over $\epem$ and $\mumu$ modes, $K^{*+}$ submodes and the low $q^2$ and high $q^2$ regions. Figure~\ref{fig:mes-low} shows the corresponding $\mes$ distributions in the low $q^2$ region. We fit the $\mes$ distributions to extract signal and background yields, $N_S$ and $N_B$ respectively. We use an ARGUS shape~\cite{ArgusShape} to describe the combinatorial background, allowing the shape parameter to float in the fits. For the signal, we use a fixed Gaussian shape unique to each final state, with mean and width determined from fits to the analogous final states in the vetoed $J/\psi$ events. We account for a small contribution from hadrons misidentified as muons by constructing a histogram pdf using $K^{(*)} h^{\pm}\mu^{\mp}$ events weighted by the probability for the $h^{\pm}$ to be misidentified as a muon. We also account for charmonium events that escape the veto, and for cross-feed contributions from misreconstructed signal events. In the entire $q^2$ region, we observe significant signal yields ($> 4 \sigma$) in each of theses modes except for $K^0 \ell^+ \ell^-$ where the significance is only $1.7\sigma$. In the low $q^2$ region significant yields ($> 4 \sigma$) are only seen in the $K^+ \ell^+ \ell^-$ and $K^{*+} \ell^+ \ell^-$ modes. The $K^0 \ell^+ \ell^-$ mode is not seen at all.

Averaging over $B^+$ and $B^0$ modes  in the low $q^2$ region we measure preliminary branching fractions  of

\begin{eqnarray}
{\cal B}(B \ra K \ell^+ \ell^-)&=&(1.8 \pm 0.4 \pm 0.08) \times 10^{-7}  \nonumber \\
{\cal B}(B \ra K^* \ell^+ \ell^-)&=&(4.3^{+1.1}_{-1.0} \pm 0.3) \times 10^{-7}
\label{eq-bf-low}
\end{eqnarray}

that are of similar size as those in the high $q^2$ region:
\begin{eqnarray}
{\cal B}(B \ra K \ell^+ \ell^-)&=&(1.4 \pm 0.4 \pm 0.07) \times 10^{-7}  \nonumber \\
{\cal B}(B \ra K^* \ell^+ \ell^-)&=&(4.2 \pm 1.0 \pm 0.3) \times 10^{-7}.
\label{eq-bf-high}
\end{eqnarray}

\begin{figure}[h]
\centering
\includegraphics[width=70mm]{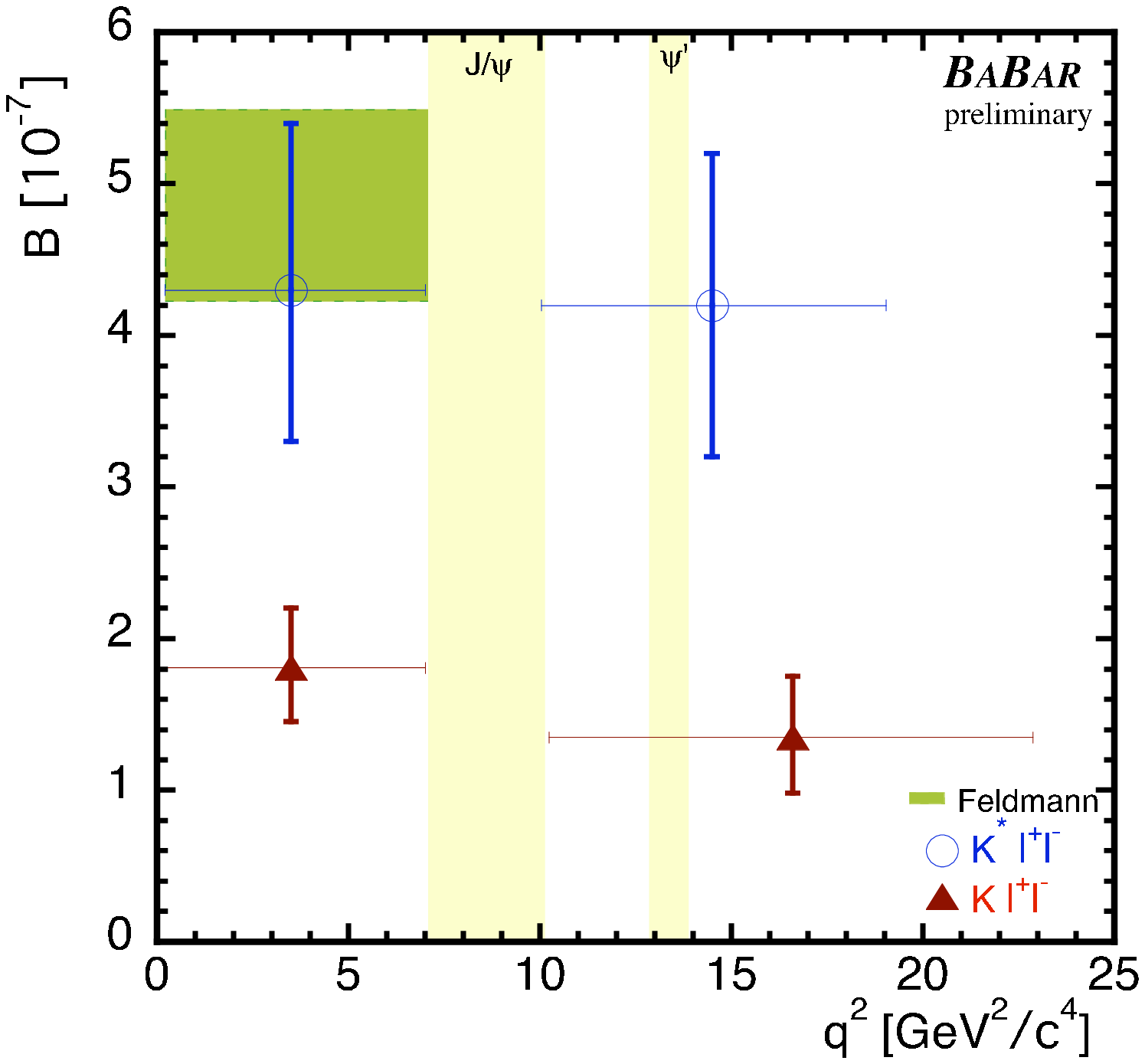}
\caption{Preliminary \babar $B \ra K \ell^+ \ell^-$ (open circles) and $B \ra K^* \ell^+ \ell^-$ (triangles) branching fraction measurements in the low $q^2$ and high $q^2$ regions. The dark-green shaded region shows the SM prediction for  $B \ra K^* \ell^+ \ell^-$.}
 \label{fig:partial-bf}
\end{figure}

Figure~\ref{fig:partial-bf} shows the \babar measurements in comparison to a SM prediction for $K^* \ell^+ \ell^-$ in the low $q^2$ region \cite{beneke05, feldmann}. The isospin averaged branching fraction is in good agreement with this prediction. 

Combining the two $q^2$ regions and correcting for the vetoed $J/\psi$ and $\psi(2S)$ regions we measure preliminary total branching fractions of

\begin{eqnarray}
{\cal B}(B \ra K \ell^+ \ell^-)&=&(3.9 \pm 0.7 \pm 0.2) \times 10^{-7}  \nonumber \\
{\cal B}(B \ra K^* \ell^+ \ell^-)&=&(11.1^{+1.9}_{-1.8} \pm 0.7) \times 10^{-7}.
\label{eq-bf-all}
\end{eqnarray}

\begin{figure}[h]
\centering
\includegraphics[width=87mm]{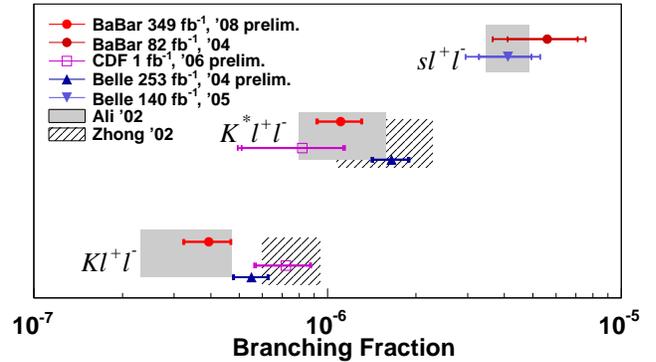}
\caption{Measured $B \ra K^{(*)} \ell^+ \ell^-$ and $B \ra X_s \ell^+ \ell^-$ branching fractions from \babar (solid points), Belle (triangles) and CDF (squares) in comparison to the SM predictions (shaded regions).}
 \label{fig:bf}
\end{figure}

Figure~\ref{fig:bf} shows recent total branching fractions measurements for $B \ra K \ell^+ \ell^-$ and
$B \ra K^* \ell^+ \ell^-$ from \babar \cite{babar06}, Belle \cite{belle2}, and CDF \cite{cdf} in comparison to two SM predictions \cite{ali, zhong}. For completeness, we have also included previous $B \ra X_s \ell^+ \ell^-$ branching fraction measurements from \babar \cite{babar1} and Belle \cite{belle3} in comparison to a SM prediction \cite{ali}. The new \babar exclusive measurements supersede the previous results \cite{babar06}. The \babar ${\cal B}(B \ra K \ell^+ \ell^-)$ measurement lies more than one standard deviation below the results from Belle and CDF but agrees well with one SM prediction \cite{ali}. 
For $B \ra K^* \ell^+ \ell^-$, the \babar measurement is in good agreement with the CDF result and both SM predictions but lies more than one standard deviation below the Belle result.

\subsection{Isospin Asymmetry Measurements}

\begin{figure}[h]
\centering
\includegraphics[width=70mm]{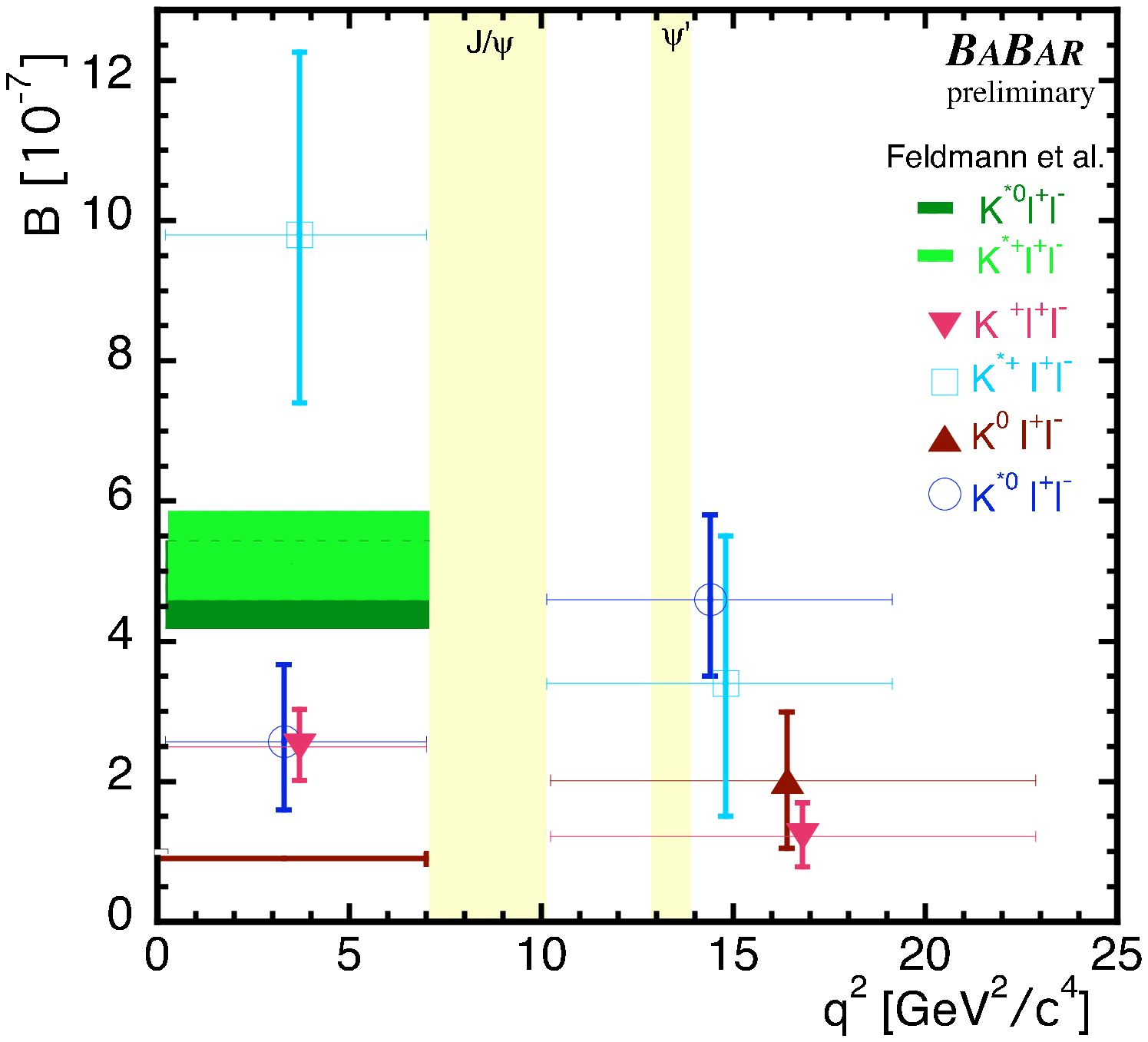}
\caption{Comparison of $B^+$ and $B^0$ branching fraction measurements for $B \ra K \ell^+ \ell^-$ and  $B \ra K^* \ell^+ \ell^-$ modes in the low $q^2$ and high $q^2$ regions, where ${\cal B}(B^+ \ra K^+ \ell^+ \ell^-)~ [{\cal B}(B^0 \ra K^0 \ell^+ \ell^-)]$ are shown by light [dark] red triangles and ${\cal B}(B^+ \ra K^{*+} \ell^+ \ell^-) ~[{\cal B}(B^0 \ra K^{*0} \ell^+ \ell^-$)] by cyan open squares [blue open circles].}
 \label{fig:bf-iso}
\end{figure}

Figure~\ref{fig:bf-iso} shows the individual \babar $B^+$ and $B^0$ branching fraction measurements in the low $q^2$ and high $q^2$ regions as well as  the SM predictions for the $K^* \ell^+ \ell^-$ modes in the low $q^2$ region~\cite{beneke05, feldmann}. While in the high $q^2$ region the isospin-related branching fractions for $B^+ \ra K^{(*)+} \ell^+ \ell^-$ and $B^0 \ra K^{(*)0} \ell^+ \ell^-$ are consistent, they differ considerably in the low $q^2$ region and deviate from the SM prediction by more than $1\sigma$. We see no signal events for $B^0 \ra K^0_S \ell^+ \ell^-$ and so we can only set a branching fraction upper limit at $90\%$ confidence level. The preliminary \babar measurements in the low $q^2$ region yield

\begin{eqnarray}
{\cal B}(B^0 \ra K^0 \ell^+ \ell^-) & < & 0.9 \times 10^{-7}~ @ 90\% ~C.L.~~~\nonumber \\
{\cal B}(B^+ \ra K^+ \ell^+ \ell^-)&=&(2.5 \pm 0.5 \pm 0.1) \times 10^{-7}  
\label{eq-bfk-low}
\end{eqnarray}
\noindent
\begin{eqnarray}
{\cal B} (B^0 \ra K^{*0} \ell^+ \ell^-)& =  &(2.6^{+1.1}_{-1.0} \pm 0.2) \times 10^{-7}  ~~ \nonumber \\
{\cal B} (B^+ \ra K^{*+} \ell^+ \ell^-)&=&(9.8^{+2.6}_{-2.4} \pm 0.6) \times 10^{-7} .
\label{eq-bfkst-low}
\end{eqnarray}

\begin{figure}[h]
\centering
\includegraphics[width=70mm]{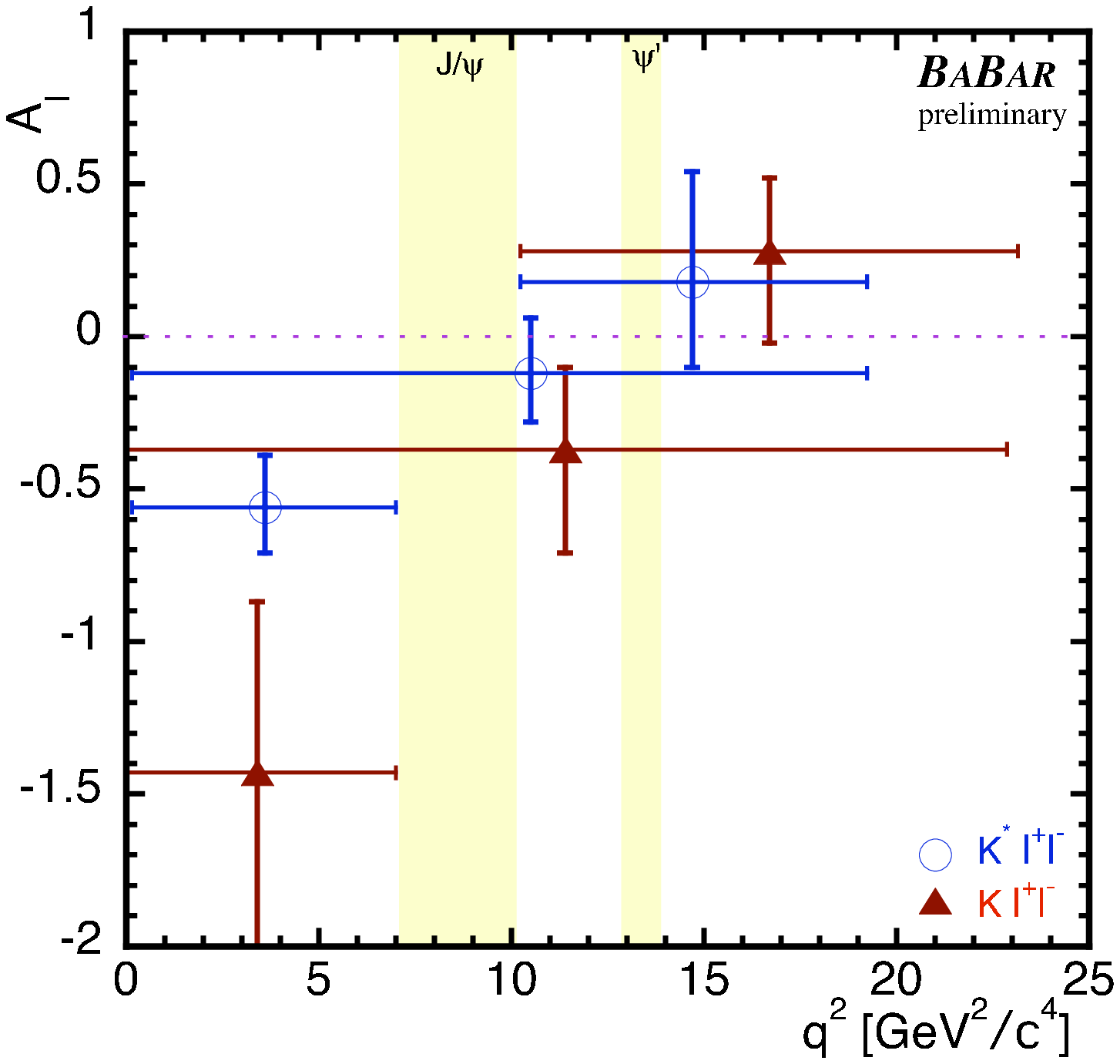}
\caption{Measured isospin asymmetries for $B \ra K \ell^+ \ell^-$ (triangles) and  $B \ra K^* \ell^+ \ell^-$ (open circles) in the low $q^2$, high $q^2$, and entire $q^2$ regions.}
 \label{fig:isospin}
\end{figure}

Table~\ref{tab:isoressys} and Figure~\ref{fig:isospin} summarize our results for the isospin asymmetries that are obtained from direct fits to the $K^{(*)0}$ and $K^{(*)+}$ data samples taking into account the different $B^0$ and $B^+$ lifetimes. Both in the high $q^2$ and entire $q^2$ regions, ${\cal A}_I$ is consistent with zero. In the low $q^2$ region, however, we observe significant isospin asymmetries.  For $K e^+ e^-$ and $K \ell^+ \ell^-$, the minimum lies in the unphysical region. For $K \mu^+ \mu^-$ due to large statistical uncertainties, the ${\cal A}_I^K$ measurement in the low $q^2$ region is consistent with both maximal isospin violation and isospin symmetry. 

\begin{figure}[h]
\centering
\includegraphics[width=65mm]{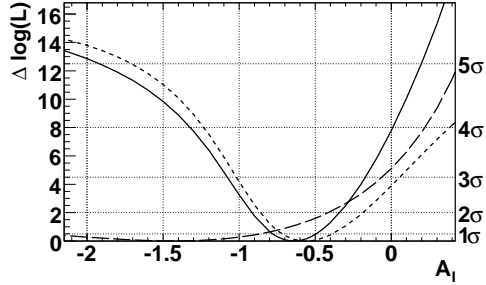}
\caption{Preliminary logarithmic likelihood curves for $B \ra K \ell^+ \ell^-$ (long-dashed curve) and $B \ra K^* \ell^+ \ell^-$ (short-dashed curve) modes using only statistical errors. The summed $\Delta \log {\cal L}$ curves are shown by the solid curve.}
 \label{fig:llike}
\end{figure}

Defining statistical significance by $N_\sigma = \sqrt{2 \Delta \log {\cal L}}$ where ${\cal L}$ is the likelihood function we determine at what values $N_\sigma$ the ${\cal A}_I = 0$ hypothesis is rejected by the data. Figure~\ref{fig:llike} shows the likelihood curves obtained from fits to the $K \ell^+ \ell^-$ and $K^* \ell^+ \ell^-$ data samples. The parabolic nature of the curves in the ${\cal A}_I > -1$ region demonstrates Gaussian nature of our fit results in the physical region. The right-side axis of Figure~\ref{fig:llike} shows purely statistical significances based on Gaussian coverage. Including systematic uncertainties, the significance of ${\cal A}_I$ in the low $q^2$ region to differ from the null hypothesis is $3.2 \sigma$ for $K \ell^+ \ell^-$ and $2.7 \sigma$ for $K^* \ell^+ \ell^-$. We have verified these confidence intervals by performing fits to MC experiments that are generated with ${\cal A}_I =0$ fixed. Using a frequentist approach we obtain results that are consistent with our above significance calculations.

\begin{figure}[h]
\centering
\includegraphics[width=70mm]{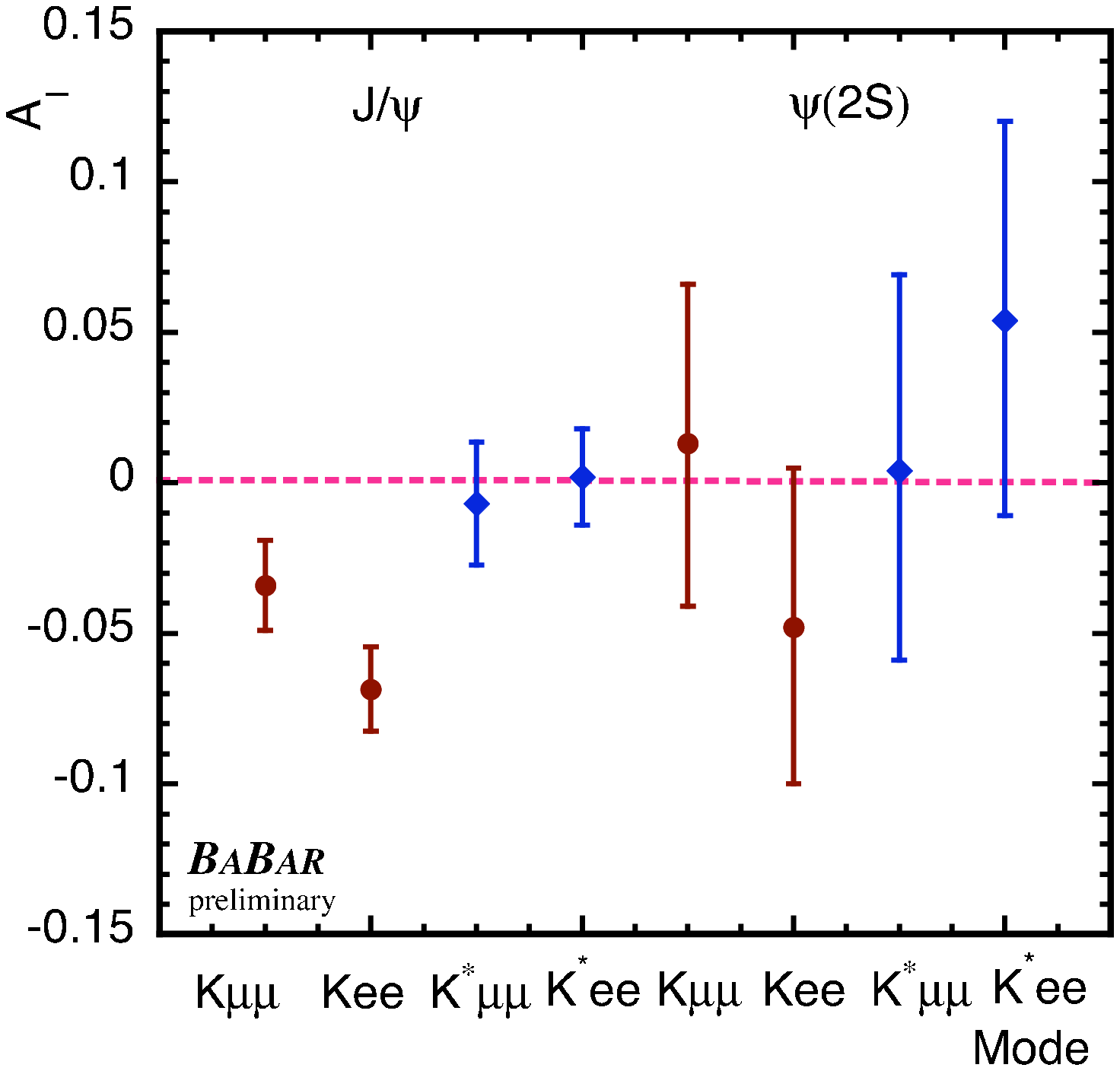}
\caption{Isospin asymmetry measurements for the charmonium control samples.}
 \label{fig:iso-cc}
\end{figure}

The highly negative ${\cal A}_I$ values for both $K \ell^+ \ell^-$ and $K^* \ell^+ \ell^-$ at low $q^2$
suggest that this asymmetry may be insensitive to the hadronic final state. Thus,  we sum the $K\ell^+ \ell^-$ and $K^* \ell^+ \ell^-$ likelihood curves as shown in Figure~\ref{fig:llike} from which we obtain ${\cal A}_I = -0.64^{+0.15}_{-0.14}\pm 0.03$ for the combined sample. Including systematic errors we find a $3.9\sigma $ significant deviation from the null hypothesis. If we Include the pole region ($ q^2 <  0.1~\rm GeV^2/c^4$) in $K^* e^+ e^-$, the isospin asymmetry is reduced to $-0.25^{+0.21}_{-0.18}\pm 0.03$. Given the large statistical error this result is consistent with the SM within two standard deviations. It is also interesting to compare the results with the isospin asymmetry measured in the $K^* \gamma$ modes. The \babar measurement of ${\cal A}_I^{K^* \gamma} =0.05\pm 0.058$ agrees well with the SM prediction \cite{babar04}. The isospin asymmetries determined for the charmonium control samples are plotted in Figure~\ref{fig:iso-cc} and are in good agreement with the SM predictions.

\begin{figure}[h]
\centering
\includegraphics[width=79mm]{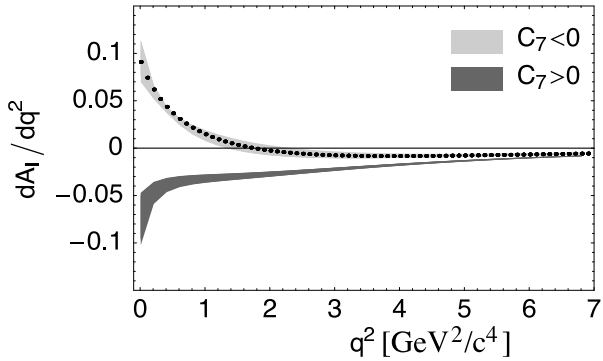}
\caption{The $q^2$ dependence of the isospin asymmetry for $B \ra K^* \ell^+\ell^-$ in the SM and for the flipped-sign $\widetilde C_7$ model \cite{feldmann02}.}
 \label{fig:ai-th}
\end{figure}

The isospin asymmetry for $K^* \ell^+ \ell^-$ modes expected in the SM is shown in Figure~\ref{fig:ai-th}. For  $q^2 \rightarrow 0$, the SM predict a positive value of $6-13\%$, which is opposite in sign to our observation. A model in which the sign of $\widetilde C_7$ is flipped provides qualitatively a better description of our results than the SM. An SM calculation of  $K^{*+}$ and $K^{*0}$ partial decay rates integrated over the low $q^2$ region yields an isospin asymmetry prediction of ${\cal A}_I = -0.005 \pm 0.02$~\cite{feldmann02, feldmann}. The \babar ${\cal A}_I^{K^*}$ result is consistent with this prediction at the $< 3\sigma $ level.

\begin{table}
\centering
\caption{Preliminary \babar measurements of isospin asymmetries in different $q^2$ regions.  Uncertainties are statistical and systematic, respectively. The last line shows $K^* e^+ e^-$ results including the $q^2 < 0.1~\rm GeV^2/c^4$ region.}
{\footnotesize
\begin{tabular}{lcccc}
Mode          & all $q^2$  & low $q^2$ & high $q^2$
\\ \hline
$K \mu^+ \mu^-$ & $0.13_{-0.37}^{+0.29}\pm0.04$   &  $-0.91_{-\infty}^{+1.2}\pm0.18$  & $0.39_{-0.46}^{+0.35}\pm0.04$ \\
$K e^+ e^-$ & $-0.73_{-0.50}^{+0.39}\pm0.04$  & $-1.41_{-0.69}^{+0.49} \pm 0.04$        & $0.21_{-0.41}^{+0.32}\pm0.03$ \\
$K \ell^+ \ell^-$ & $-0.37_{-0.34}^{+0.27}\pm0.04$  & $-1.43_{-0.85}^{+0.56} \pm 0.05$        & $0.28_{-0.30}^{+0.24}\pm0.03$ \\
$K^* \mu^+ \mu^-$  & $-0.00_{-0.26}^{+0.36}\pm0.05$  & $-0.26_{-0.34}^{+0.50} \pm0.05$         &$-0.08_{-0.27}^{+0.37}\pm0.05$ \\
$K^* e^+ e^-$  & $-0.20_{-0.20}^{+0.22}\pm0.03$  & $-0.66_{-0.17}^{+0.19} \pm0.02$         & $0.32_{-0.45}^{+0.75}\pm0.03$ \\
$K^* \ell^+ \ell^-$  & $-0.12_{-0.16}^{+0.18}\pm0.04$  & $-0.56_{-0.15}^{+0.17} \pm0.03$         & $0.18_{-0.28}^{+0.36}\pm0.04$ \\ \hline
$K^* e^+ e^-$  & $-0.27_{-0.18}^{+0.21}\pm 0.03$ & $-0.25_{-0.18}^{+0.20}\pm0.03$          & --- &   \\
\hline \hline                        
\end{tabular}
}
\label{tab:isoressys}
\end{table}

\subsection{Direct \CP Violation Measurements}

We extract the \CP asymmetries by performing fits to the split $B$ and $\Bbar$ datasets in charge-conjugate final states using all modes except $B \ra K^0_S \ell^+ \ell^-$. We assume a common background ARGUS shape parameter $\xi$. The results for ${\cal A}_{CP}$ are summarized in Table \ref{tab:acpressys}. We observe \CP asymmetries that are consistent with zero as expected in the SM.

\begin{table}
\centering
\caption{Preliminary $A_{CP}$ results. The uncertainties are statistical and systematic, respectively.}
\label{tab:acpressys}
{\footnotesize
\begin{tabular}{lccc}
Mode & all $q^2$ & low $q^2$ & high $q^2$
\\ \hline
$K^+ \ell^+ \ell^-t$ & $-0.18_{-0.18}^{+0.18}\pm0.01$ & $-0.18_{-0.19}^{+0.19}\pm0.01$ & $-0.09_{-0.39}^{+0.36}\pm0.02$   \\
$K^{*0} \ell^+ \ell^-$ & $0.02_{-0.20}^{+0.20}\pm0.02$ & $-0.23_{-0.38}^{+0.38}\pm0.02$ & $0.17_{-0.24}^{+0.24}\pm0.02$  \\
$K^{*+} \ell^+ \ell^-$ & $0.01_{-0.24}^{+0.26}\pm0.02$ & $0.10_{-0.24}^{+0.25}\pm0.02$ & $-0.18_{-0.55}^{+0.45}\pm0.04$ \\
$K^* \ell^+ \ell^-$ & $0.01_{-0.15}^{+0.16}\pm0.01$ & $0.01_{-0.20}^{+0.21}\pm0.01$ & $0.09_{-0.21}^{+0.21}\pm0.02$    \\
\hline
\end{tabular}
}
\end{table}

\subsection{Tests of Lepton Flavor Asymmetries}

\begin{figure}[h]
\centering
\includegraphics[width=80mm]{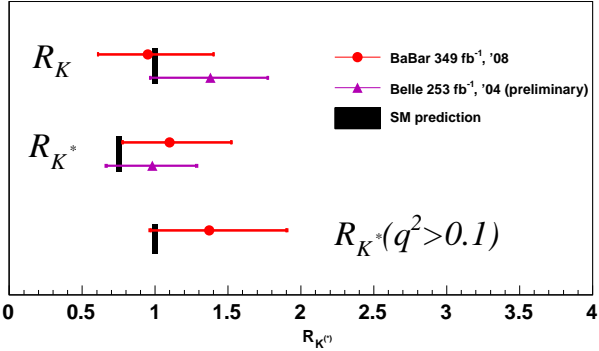}
\caption{Preliminary measurements of ${\cal R}_K$ and ${\cal R}_{K^*}$  from \babar (points) and Belle (triangles) in
comparison to the SM predictions (black bars).}
 \label{fig:rk}
\end{figure}

Table~\ref{tab:emuressys} shows the \babar results for the lepton flavor ratios ${\cal R}_{K}$ and ${\cal R}_{K^*}$ both including and excluding events with $q^2<0.1 \rm GeV^2/c^4$. The most significant deviation from the SM expectations is found in the low $q^2$ region, where ${\cal R}_{K}=0.40_{-0.23}^{+0.30}\pm 0.02$ lies $\sim 2\sigma$ below the SM prediction of one. Figure~\ref{fig:rk} shows ${\cal R}_K$ and ${\cal R}_{K^*}$ measurements from \babar and Belle~\cite{belle3} for the entire $q^2$ region in comparison to the SM predictions~\cite{ali}. At the present level of precision, all results are consistent with lepton flavor universality.

\begin{table}
\caption{Preliminary lepton flavor ratio results. The uncertainties are statistical and systematic, respectively.}
\centering
\begin{tabular}{ccc}
$q^2$ Region  & $R_{K^*}$                      & $R_{K}$
\\ \hline
all                       & $1.37_{-0.40}^{+0.53}\pm 0.09$ & $0.96_{-0.34}^{+0.44}\pm 0.05$ \\
$\mathrm{all}+ q^2 < 0.1 $ & $1.10_{-0.32}^{+0.42}\pm 0.07$ & ---                            \\
low                       & $1.01_{-0.44}^{+0.58}\pm 0.08$ & $0.40_{-0.23}^{+0.30}\pm 0.02$ \\
$\mathrm{low}+ q^2 < 0.1 $ & $0.56_{-0.23}^{+0.29}\pm 0.04$ & ---                            \\
high                      & $2.15_{-0.78}^{+1.42}\pm 0.15$ & $1.06_{-0.51}^{+0.81}\pm 0.06$ \\
\hline
\end{tabular}
\label{tab:emuressys}
\end{table}

\section{Angular Analysis of $ B \rightarrow K^* \ell^+ \ell^-$}

The data selection for the angular analysis is performed in a similar way as that for the decay rates, except that some selection criteria are tightened to minimize backgrounds~\cite{run5afb}. Using 384~million $B \bar B$ events we select the same six $K^* \ell^+ \ell^-$ final states as in section~\ref{kll}. We measure ${\cal F}_L$ and ${ \cal A}_{FB}$ in the low $q^2$ and the high $q^2$ regions, where the low $q^2$ region is reduced to $0.1 <q^2 < 6.25~ \rm GeV^2/c^4$  to ascertain that leakage from $J/\psi K^*$ background
into the final data sample is rather small. The final reconstruction efficiencies for signal events vary from $1.5\%$ for $K^+ \pi^0 \mu^+ \mu^-$ in the low $q^2$ region to $12.6\%$ for $K^+\pi^- e^+e^-$ in the high $q^2$ region.

Due to the small event samples, we fit the data to the one-dimensional angular distributions in Eq~\ref{eq-theta}. Note, that ${\cal F}_L$ is constrained by $W(\cos \theta_K)$ and for fixed ${\cal F}_L$, ${\cal A}_{FB}$ is constrained by $W(\cos \theta_\ell)$. For each $q^2$ region, we combine events from all six final states and perform three successive unbinned maximum likelihood fits. First, we fit the $m_{ES}$ distributions in the region $m_{ES} > 5.2~\rm GeV/c^2$ to obtain the number of signal ($N_S$) and background ($N_B$) yields using a Gaussian pdf for signal with mean and width determined from the vetoed charmonium sample and an Argus shape for the combinatorial background. We account for a small contribution from hadrons misidentified as muons, for misreconstructed signal events and for charmonium events that escape the veto. 

In the second fit to the $\cos \theta_K$ distribution for events with $m_{ES} > 5.27~\rm GeV/c^2$ we extract ${\cal F}_L$, where the normalization for signal and background events is taken from the first fit to $m_{ES}$. In the third fit to the $\cos \theta_\ell$ distribution for events with $m_{ES} > 5.27~ GeV/c^2$ we extract ${\cal A}_{FB}$ for ${\cal F}_L$ fixed to the result from the previous fit and the normalization determined from the fit to $m_{ES}$. We model the $\cos \theta_K$ and $\cos \theta_\ell$ shapes of the combinatorial background using $ e^+ e^-$ and $\mu^+ \mu^-$ events, as well as lepton flavor violating $e^\pm \mu^\mp$  events  in the $ 5.20 < m_{ES}  < 5.27~\rm  GeV/c^2$ sideband. The signal distribution is convolved with the detector acceptance as a function of $\cos \theta_K$ and $\cos \theta_\ell$, respectively. The correlated leptons from $B \ra D^{(*)} \ell \nu$, $D \ra K^{(*)} \ell \nu$ give rise to a peak in the combinatorial background at $\cos \theta_\ell > 0.7$ which varies as a function of $m_{ES}$. We consider this variation in our study of systematic errors.

\begin{figure}[h]
\centering
\includegraphics[width=77mm]{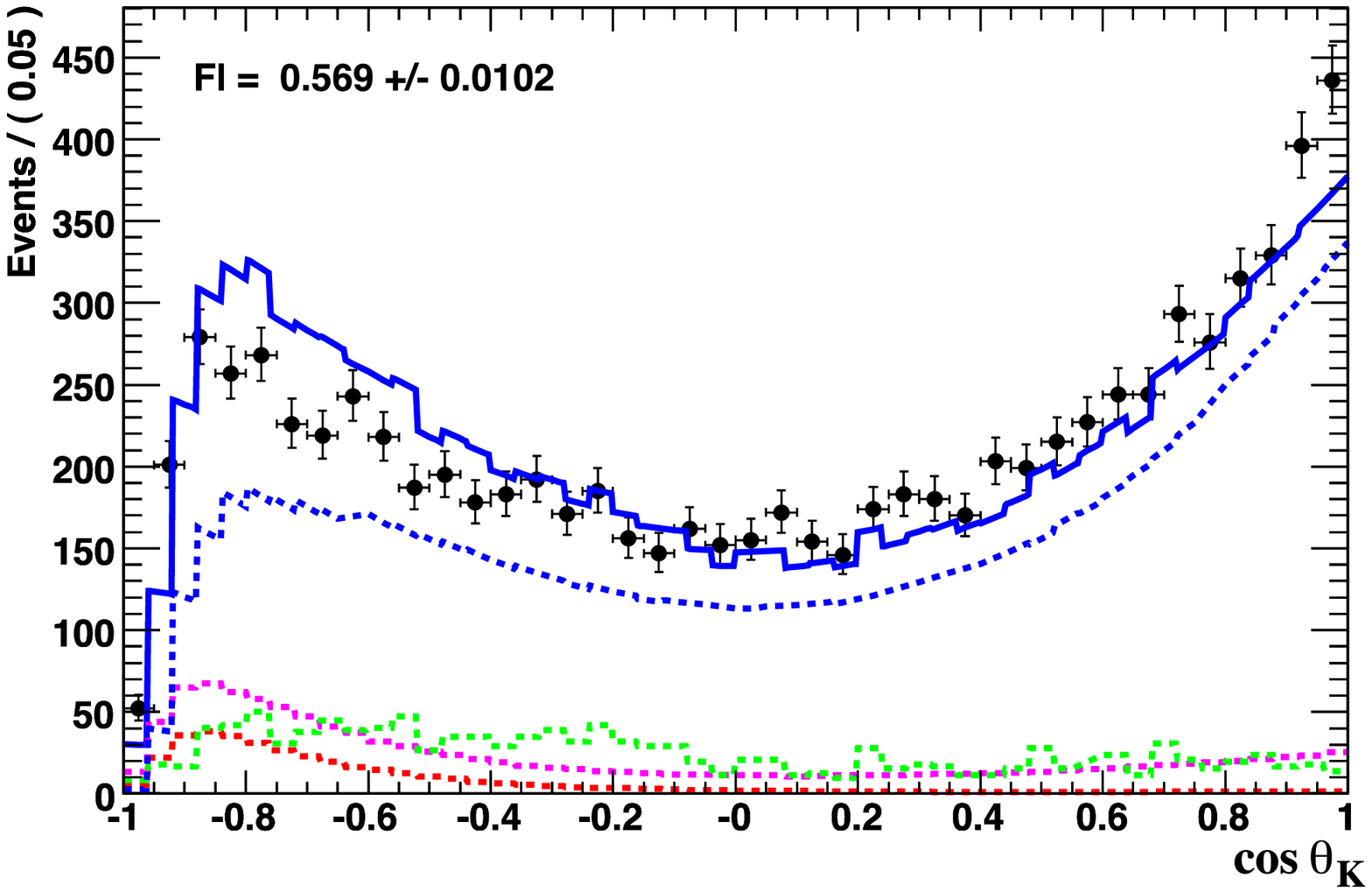}
\includegraphics[width=77mm]{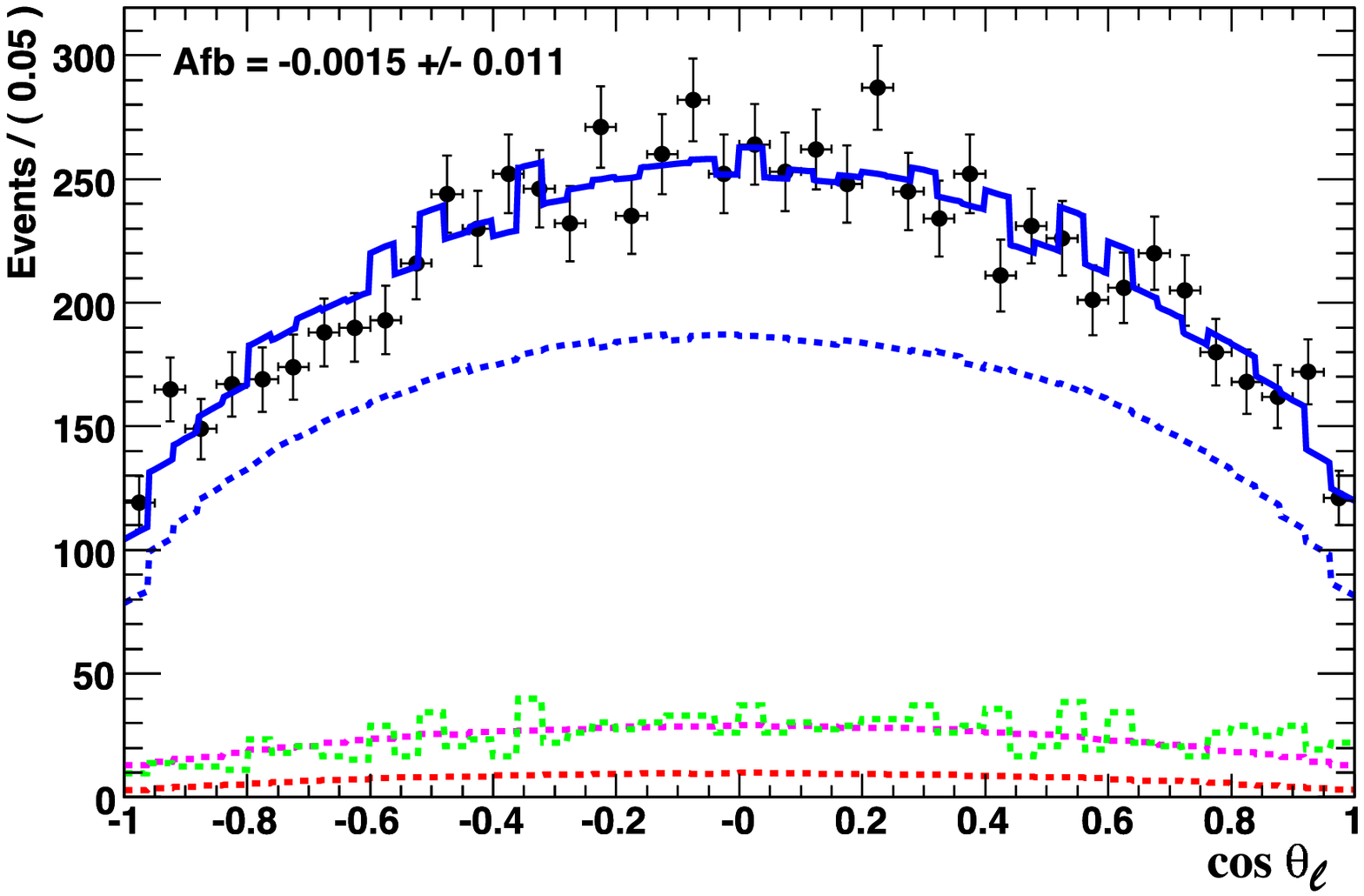}
\caption{\babar measurements of the $\cos \theta_K$ (top) and $\cos \theta_\ell$ (bottom) distributions (points with error bars) for the $B \ra J/\psi (\ra \ell^+ \ell^-) K^* $ sample for all six modes combined with fit results superimposed, total fit (solid blue line), signal contribution (blue dots), combinatorial background (green-dahed line), self cross-feed  (magenta dashed line) and feed-across background (red dashed line).}
 \label{fig:angle-cc}
\end{figure}

\begin{figure}[h]
\centering
\includegraphics[width=70mm]{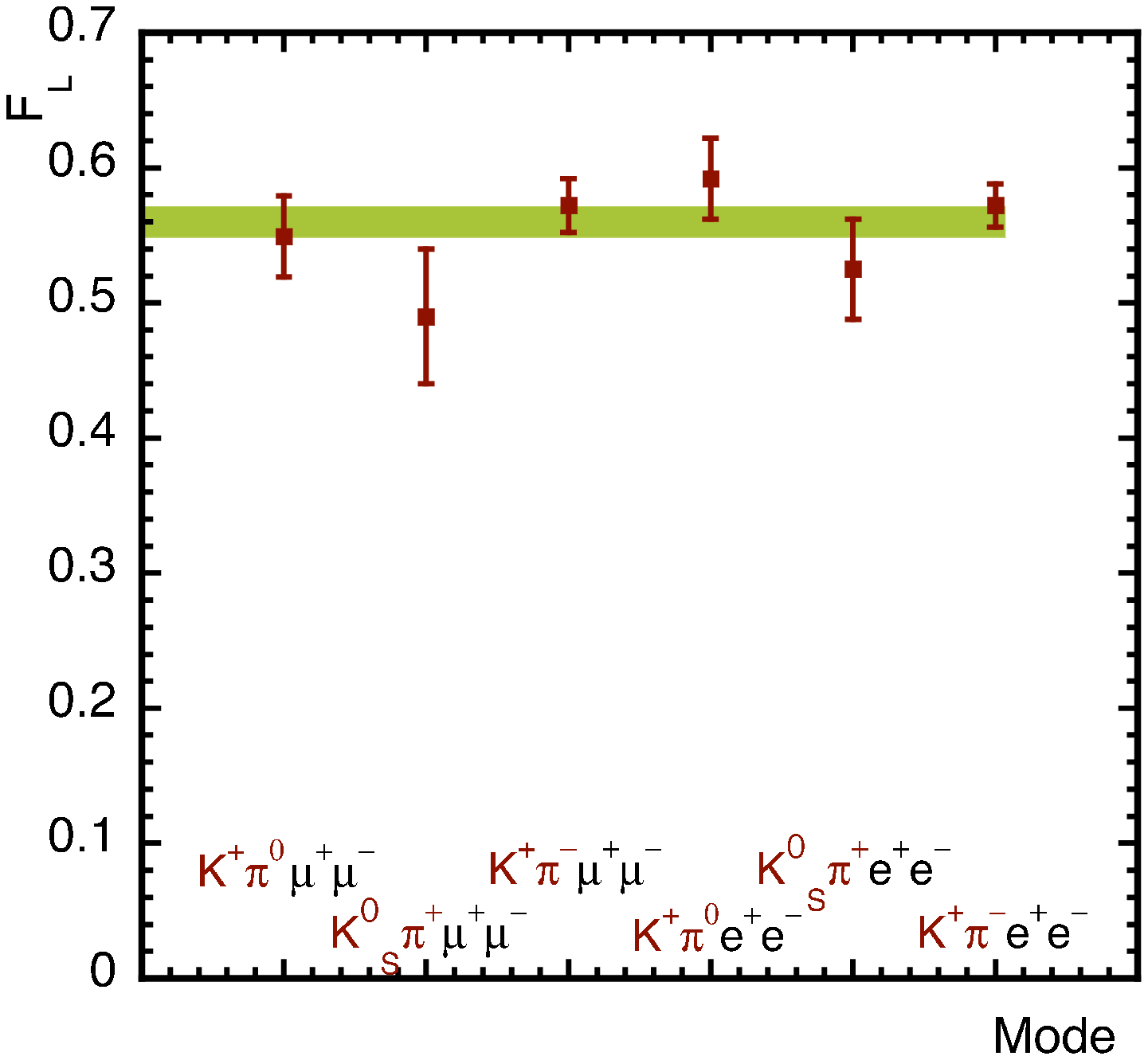}
\caption{\babar measurements of ${\cal F}_L$ values for six $J/\psi K^*$ control samples. The green bar represents the result of a recently published \babar analysis.}
 \label{fig:fl-cc}
\end{figure}

\begin{figure}[h]
\centering
\includegraphics[width=70mm]{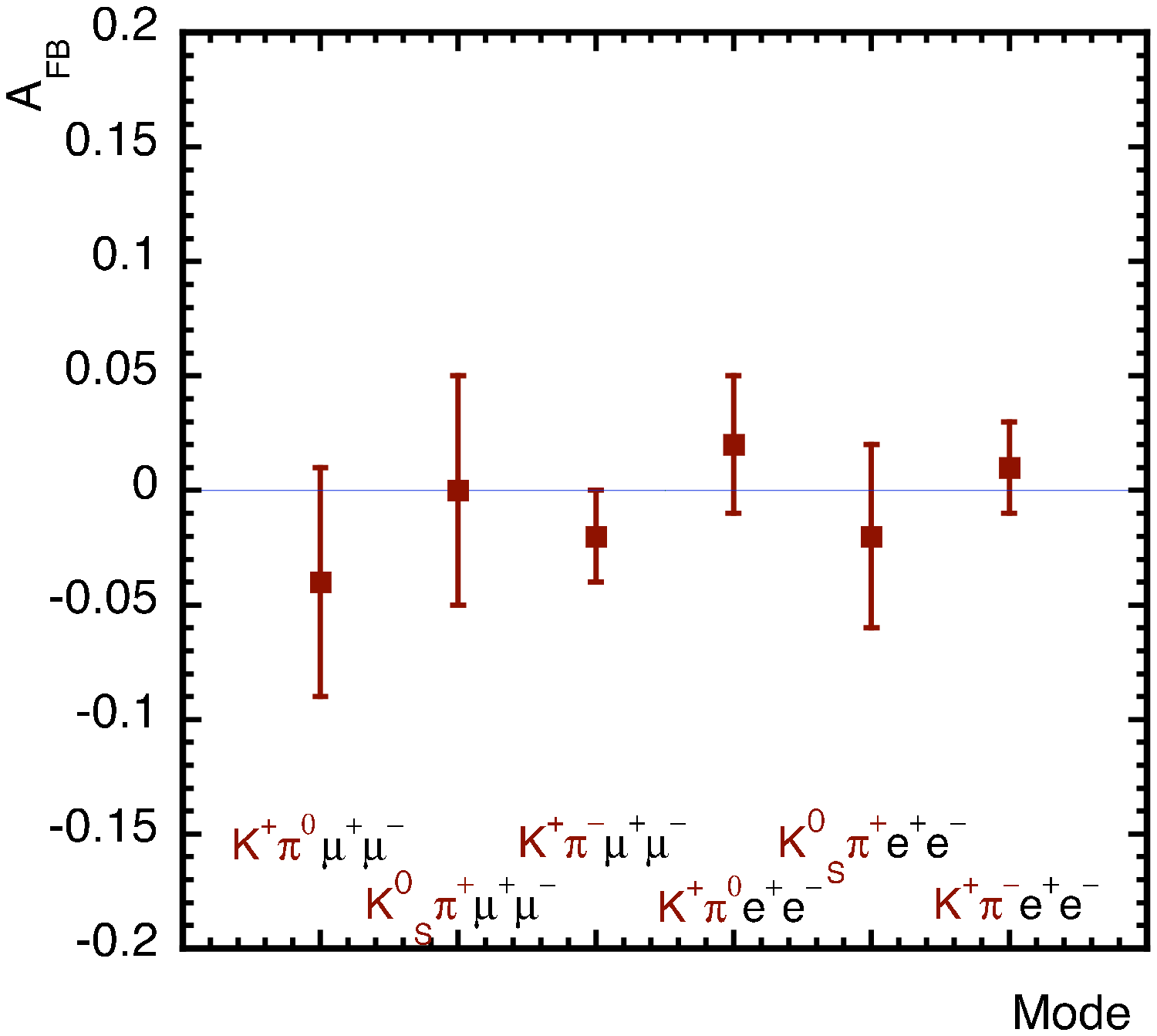}
\caption{\babar measurements of ${\cal A}_{FB}$ for six $J/\psi K^*$ control samples.}
 \label{fig:afb-cc}
\end{figure}

We test our fit methodology by using the large samples of vetoed charmonium events. Figure~\ref{fig:angle-cc} shows the $\cos \theta_K$ and $\cos \theta_\ell$ distributions for all $B \ra J/\psi K^*$ events. We extract  ${\cal F}_L =0.569\pm 0.010$ and ${\cal A}_{FB} =-0.001 \pm 0.011$. While ${\cal A}_{FB} $ is consistent with zero as expected in the SM, ${\cal F}_L$ agrees with a recent \babar measurement yielding ${\cal F}_L =0.56 \pm 0.01$ \cite{babar07}. The ${\cal F}_L$ and ${\cal A}_{FB}$ results for the individual $J/\psi K^*$ submodes are shown in Figures~\ref{fig:fl-cc} and \ref{fig:afb-cc}. 
We also fit the $m_{ES}$ and $\cos \theta_\ell$ distributions of the $K^+ \ell^+ \ell^-$ decays and find ${\cal A}_{FB}$ consistent with zero as expected.  We have performed fits using signal events generated with different values of the Wilson coefficients covering the allowed ranges of ${\cal F}_L$ and ${\cal A}_{FB}$.
We find no bias in extracting  ${\cal F}_L$ and ${\cal A}_{FB}$ from the fits.

\begin{figure}[h]
\centering
\includegraphics[width=85mm]{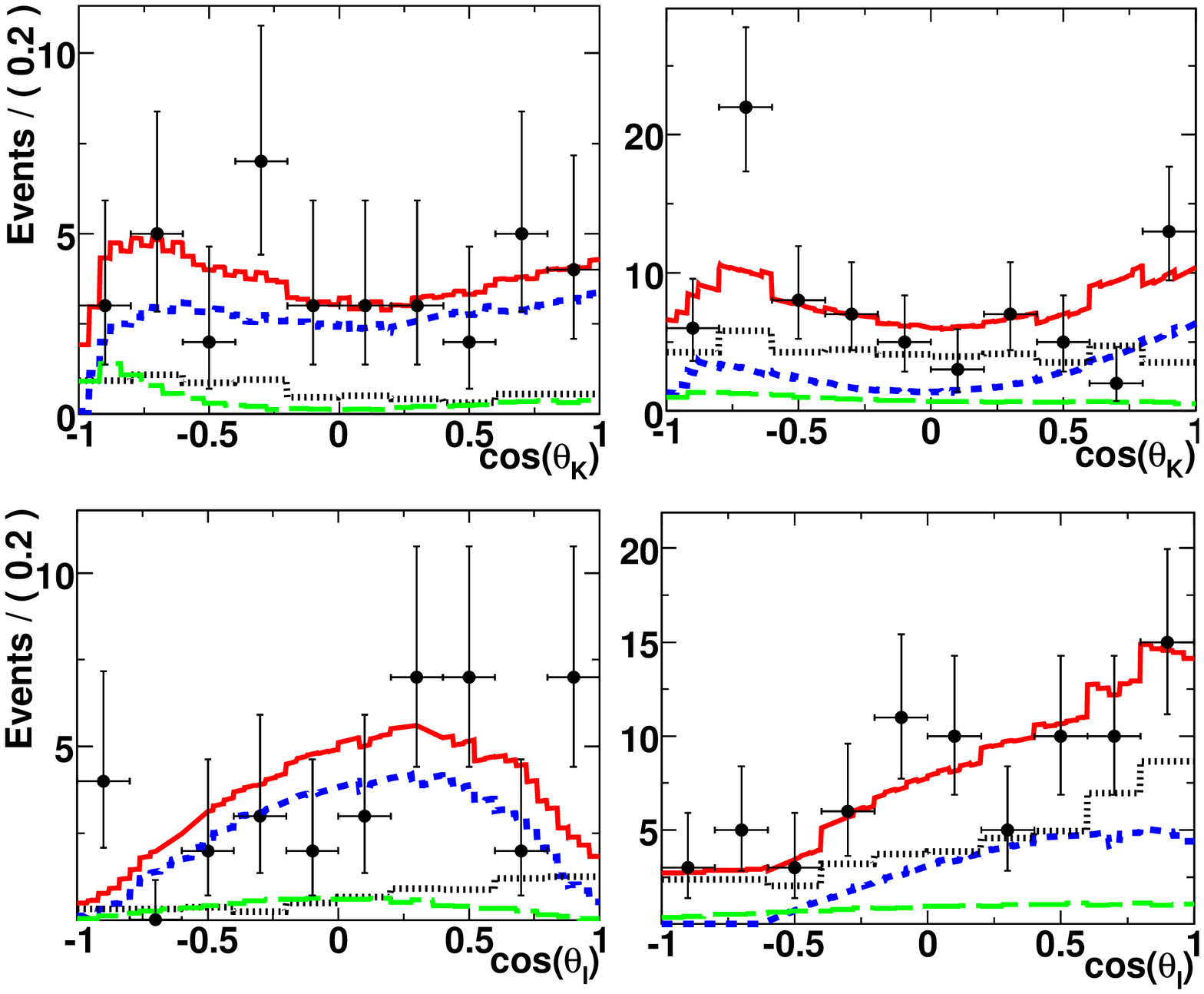}
\caption{\babar measurements of the $\cos \theta_K$ (top) and $\cos \theta_\ell$ (bottom) distributions (points with error bars) for the six combined $K^* \ell^+ \ell^-$ signal modes in the low $q^2$ (left) and high $q^2$ (right) regions with fit results superimposed, total fit (red solid histogram), signal contribution (blue dashes), combinatorial background (black dots) and peaking backgrounds (green long dashes).}
 \label{fig:angle-data}
\end{figure}

Figure~\ref{fig:angle-data} shows the $\cos \theta_K$ and $\cos \theta_\ell$ distributions for the six $K^* \ell^+ \ell^-$ modes combined in the low $q^2$ and high $q^2$ regions. The fits include pdf shapes of the signal, combinatorial background, peaking backgrounds, and backgrounds from misidentified muons. In the low $q^2$ and high $q^2$ regions we measure

\begin{eqnarray}
{\cal F}_L^{low} &=& 0.35\pm 0.16 \pm 0.04  \nonumber \\
{\cal F}_L^{high} &=& 0.71^{+0.20}_{-0.22}\pm 0.05
\label{eq-fl} 
\end{eqnarray}
and
\begin{eqnarray}
{\cal A}_{FB}^{low} &=& 0.24^{+0.18}_{-0.23} \pm 0.06  \nonumber \\
{\cal A}_{FB}^{high} &=& 0.76^{+0.52}_{-0.32} \pm 0.07.
\label{eq-afb}
\end{eqnarray}

\begin{figure}[h]
\centering
\includegraphics[width=72mm]{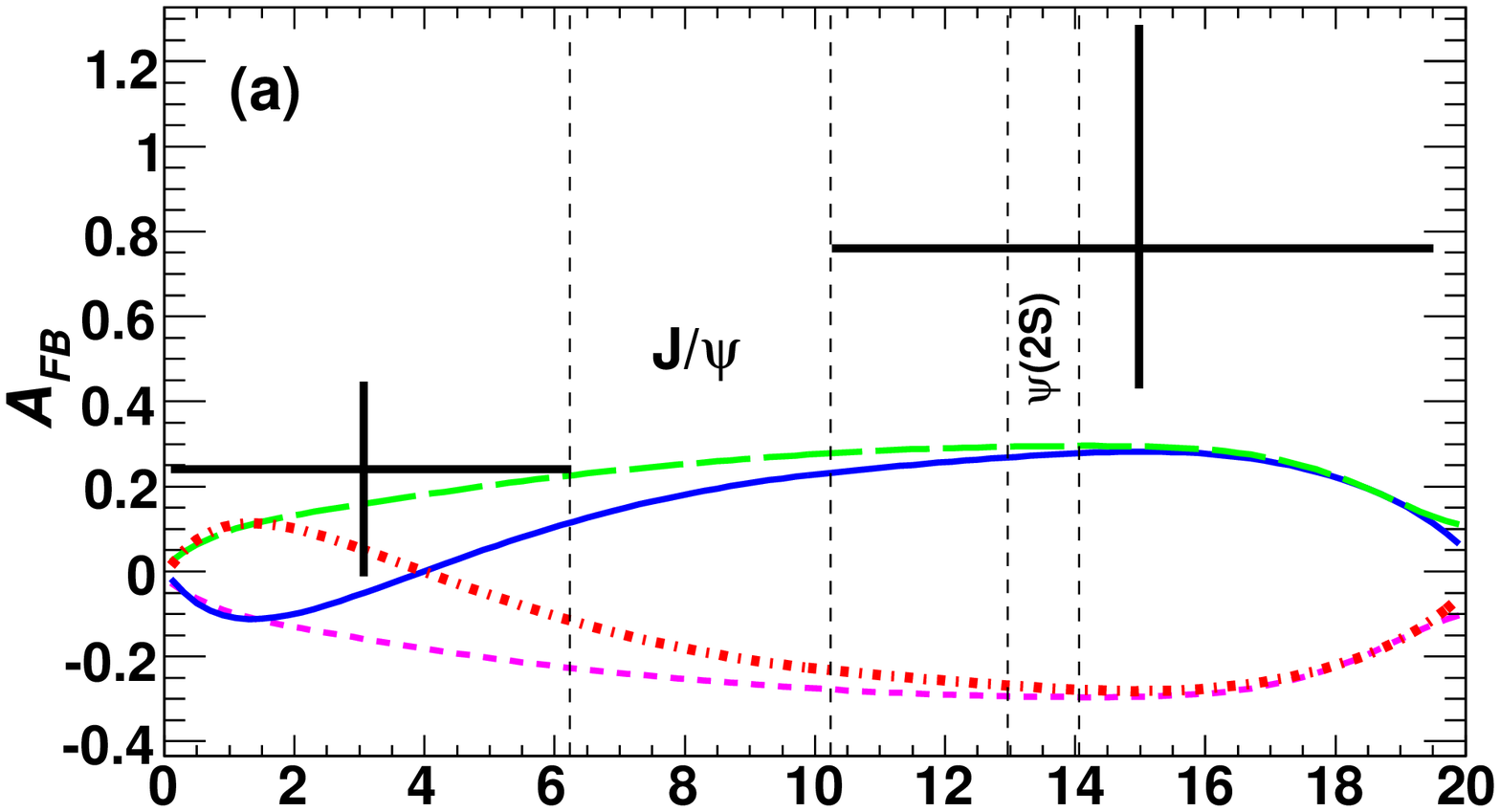}
\includegraphics[width=70mm]{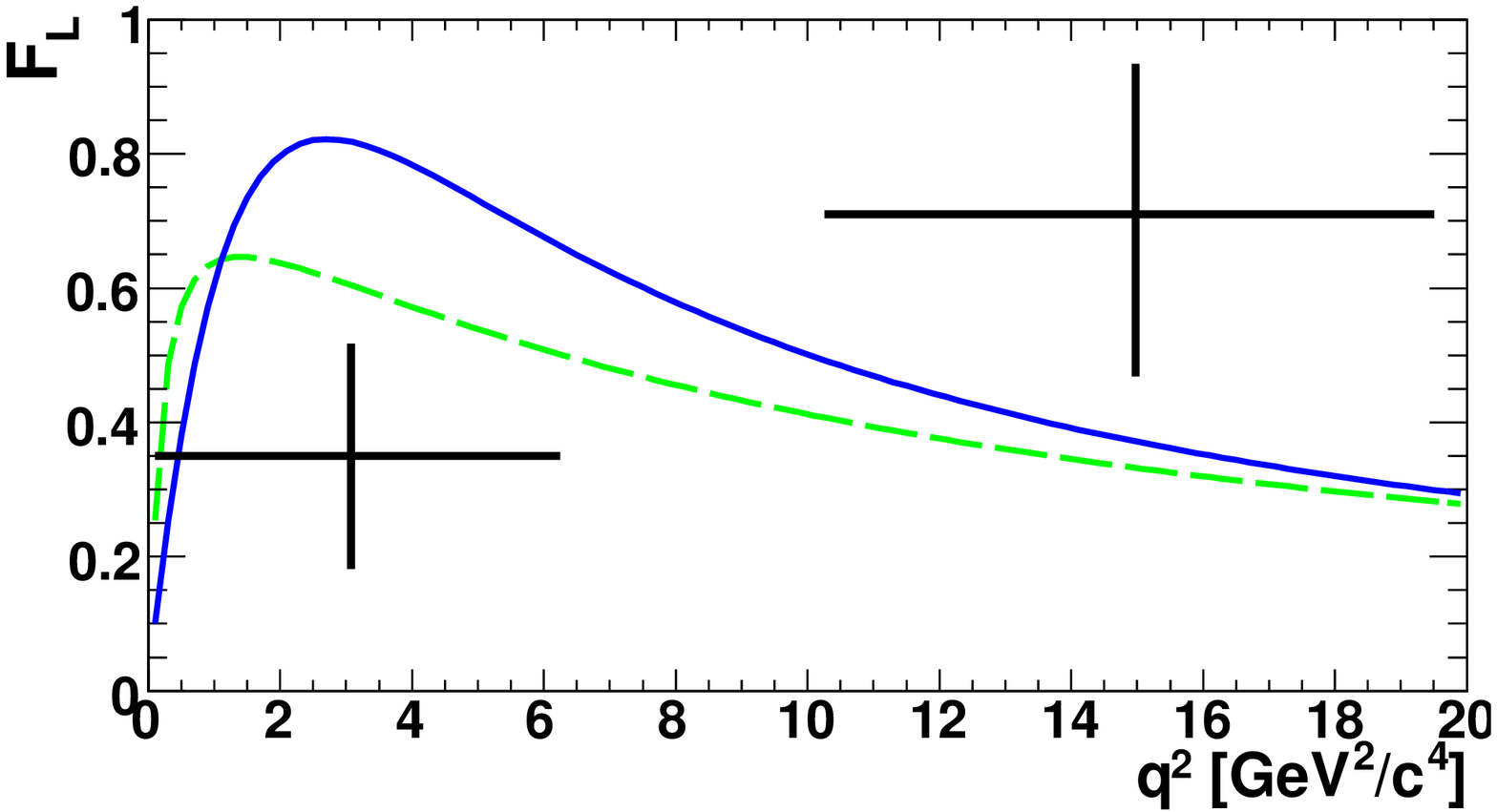}
\caption{\babar measurements of ${\cal A}_{FB}$ (top) and ${\cal F}_L$ (bottom) in the low $q^2$ and high $q^2$ regions (points with error bars). For comparison, predictions are shown for the SM (solid curve), the flipped-sign $\widetilde C_7$ model (green dashed curve),  the flipped-sign $\widetilde C_9 \widetilde C_{10}$ model (magenta dotted curve) and the mirror image model of the SM (red dash-dotted curve). The solid (dashed) lines in the ${\cal F}_L$ plot show the SM (flipped $\widetilde C_7$ model) predictions after integrating over the low $q^2$ and high $q^2$ regions separately.}
 \label{fig:afb-fl}
\end{figure}

\begin{figure}[h]
\centering
\includegraphics[width=80mm]{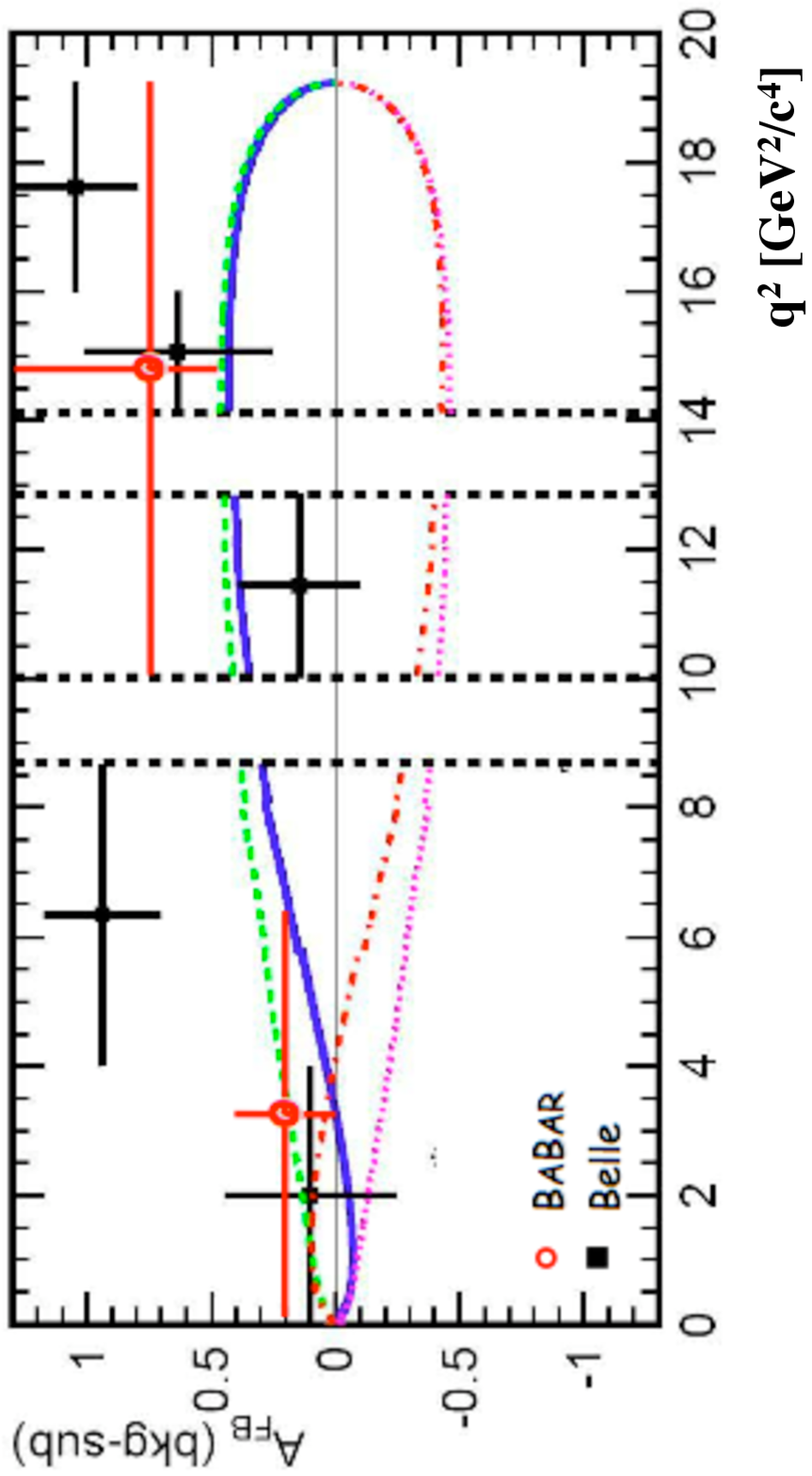}
\caption{Comparison of \babar (open circles) and Belle (solid points) results for ${\cal A}_{FB}$ as a function of $q^2$. }
 \label{fig:afb-comp}
\end{figure}

Figures~\ref{fig:afb-fl} show these results in comparison to the SM prediction and expectations of the three other models introduced in chapter~\ref{theory}. Though consistent with the SM prediction, the results for ${\cal F}_L$ and ${\cal A}_{FB}$ favor the flipped-sign $\widetilde C_7$ model \cite{hou}. Figure~\ref{fig:afb-comp} shows our  ${\cal A}_{FB}$ results in comparison to the present Belle results \cite{belle}. In the Belle analysis the low $q^2$ region is divided into two bins where the high bin extends to $8.7~ \rm  GeV^2/c^4$. The high $q^2$ region is divided into a bin between the $J/\psi$ and $\psi(2S)$ resonances and two bins above the $\psi(2S)$. The results of both experiments are in good agreement and the Belle results also favor the flipped-sign  $\widetilde C_7$ model. The large values of ${\cal A}_{FB}$ in the high $q^2$-region disfavor the wrong-sign $\widetilde C_9 \widetilde C_{10}$ model at the $ > 3 \sigma$ level. Obviously, we need to perform a model-independent global fit to measurements of ${\cal A}_{FB}$, ${\cal F}_L$ and other observables using data from \babar and Belle to look for significant discrepancies to the SM predictions and to extract non-SM contributions in the effective Wilson coefficients $\widetilde C_7$, $\widetilde C_9$ and $\widetilde C_{10}$. This, however, may require more precise results than are presently available. In addition, it would be useful to include measurements of other electroweak penguin decays  such as  $B \ra X_s \ell^+ \ell^-$ in the fit.

\section{Search for $B \ra \pi \ell^+ \ell^-$}

\begin{figure}[h]
\centering
\includegraphics[width=70mm]{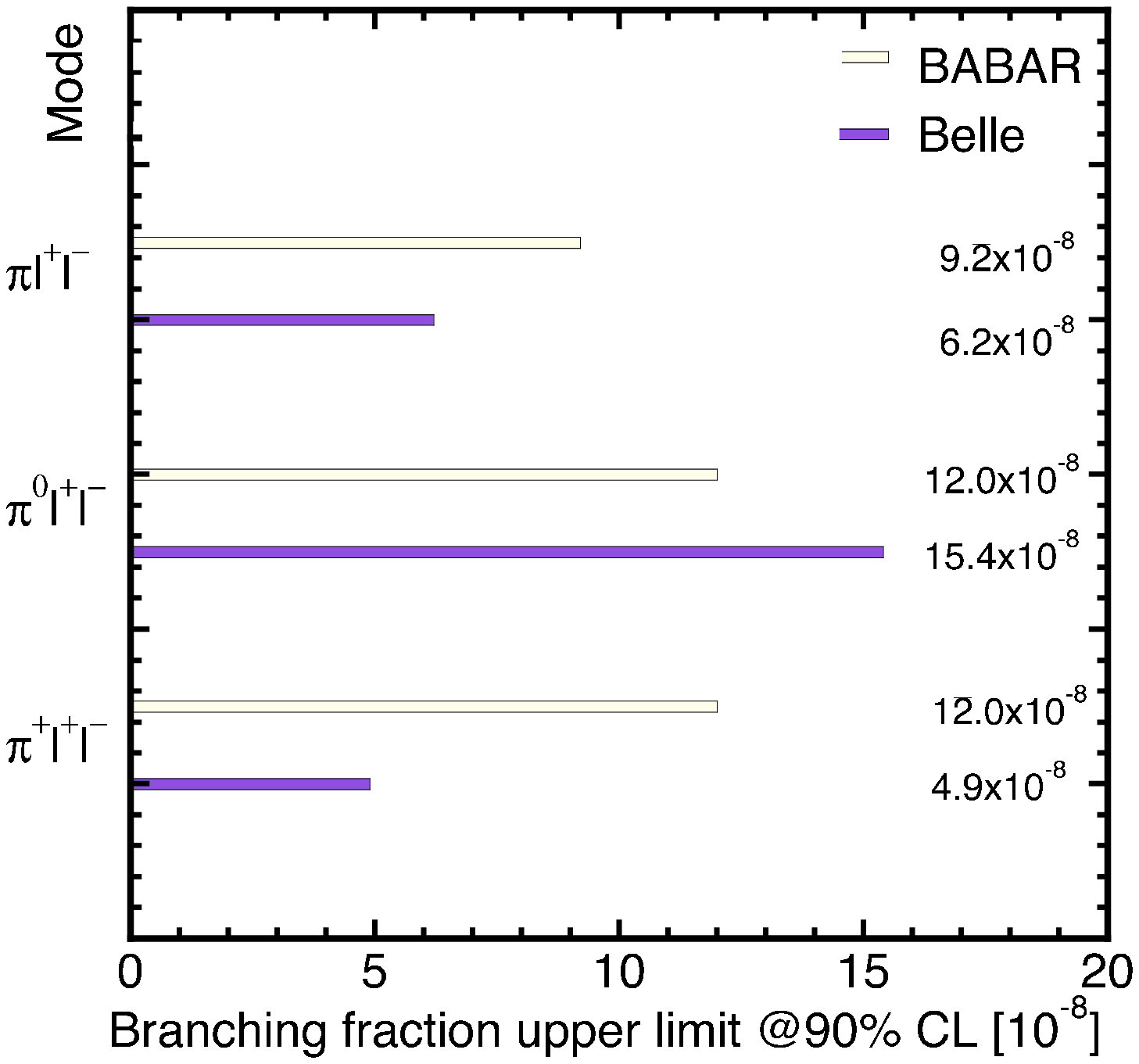}
\caption{Comparison of branching fraction upper limits at $90\% $ C.L. of $B \ra \pi \ell^+ \ell^-$ modes from \babar (yellow bars) and Belle (purple bars).}
 \label{fig:pill}
\end{figure}

The exclusive decay $B \ra \pi \ell^+ \ell^-$ is a $b \ra d$ transition and is suppressed with respect to $B \ra K \ell^+ \ell^-$ by $|V_{td}/V_{ts}|^2$. The branching fraction expected in the SM  is  ${\cal B}(B^+ \ra \pi^+ \ell^+ \ell^-)\simeq 3.3 \times 10^{-8}$ \cite{buchalla, ali3}. Belle has updated a search for $B \ra \pi \ell^+ \ell^-$ for both $\pi^\pm$ and $\pi^0$ in the recoil of an $e^+ e^-$ or $\mu^+ \mu^-$ pair using 657 million $B \bar B$ events \cite{belle08}. Since the main background originates from $q \bar q$ continuum events ($ q=u,~d,~s,~c)$, Belle forms a Fisher discriminant from 16 shape variables. They combine the Fisher discriminant together with the $B$ vertex separation and the $\cos \theta_B$ distribution into a likelihood ratio, where $\theta_B$ is the angle of the $B$ meson in the $\Upsilon(4S)$ rest frame with respect to the beam axis. After removing events in the vicinity of the $J/\psi$ and $\psi(2S)$ resonance regions they perform an unbinned maximum likelihood fit in the $\Delta E- m_{ES}$ plane. Since they observe no signal events, they determine branching fraction upper limits at $90\%$ confidence level. The Belle limits are plotted in Figure~\ref{fig:pill} in comparison to \babar limits \cite{babar-pill} that used 230 million $B \bar B$ events. Belle has set the lowest limits for $\pi^+ \ell^+ \ell^-$ and for the combined $\pi \ell^+ \ell^-$ modes, whereas \babar has set the lowest limit for $\pi^0 \ell^+ \ell^-$. The Belle result of ${\cal B}(B^+ \ra \pi^+ \ell^+ \ell^-) < 4.9 \times 10^{-8} ~@ 90\% ~CL$ lies just a factor of $\sim 1.5$ above the SM prediction.

\section{Search for $B \ra K^{(*)} \nu \bar \nu$}

The decays $B \ra K^{(*)} \nu \bar \nu$ proceed through the Z-penguin and weak-box diagrams where $\ell \ell$ is now a $\nu \bar \nu$ pair. In the SM the branching fractions are predicted to be \cite{buchalla, ali3}:                                  

\begin{eqnarray}
{\cal B}(B \ra K^\pm \nu \bar \nu) &=& (3.8^{+1.2}_{-0.6}) \times 10^{- 6} \nonumber \\
{\cal B}(B \ra K^* \nu \bar \nu) &=& (13^{+4}_{-3}) \times 10^{- 6}
\label{eq-knunu}
\end{eqnarray}

New physics may modify these predictions. Contributions from new loop and box-diagrams with new particles in the loop (see Figure~\ref{fig:np}) may interfere constructively or destructively with the signal modes yielding enhanced or reduced branching fractions.

\begin{figure}[h]
\centering
\includegraphics[width=80mm]{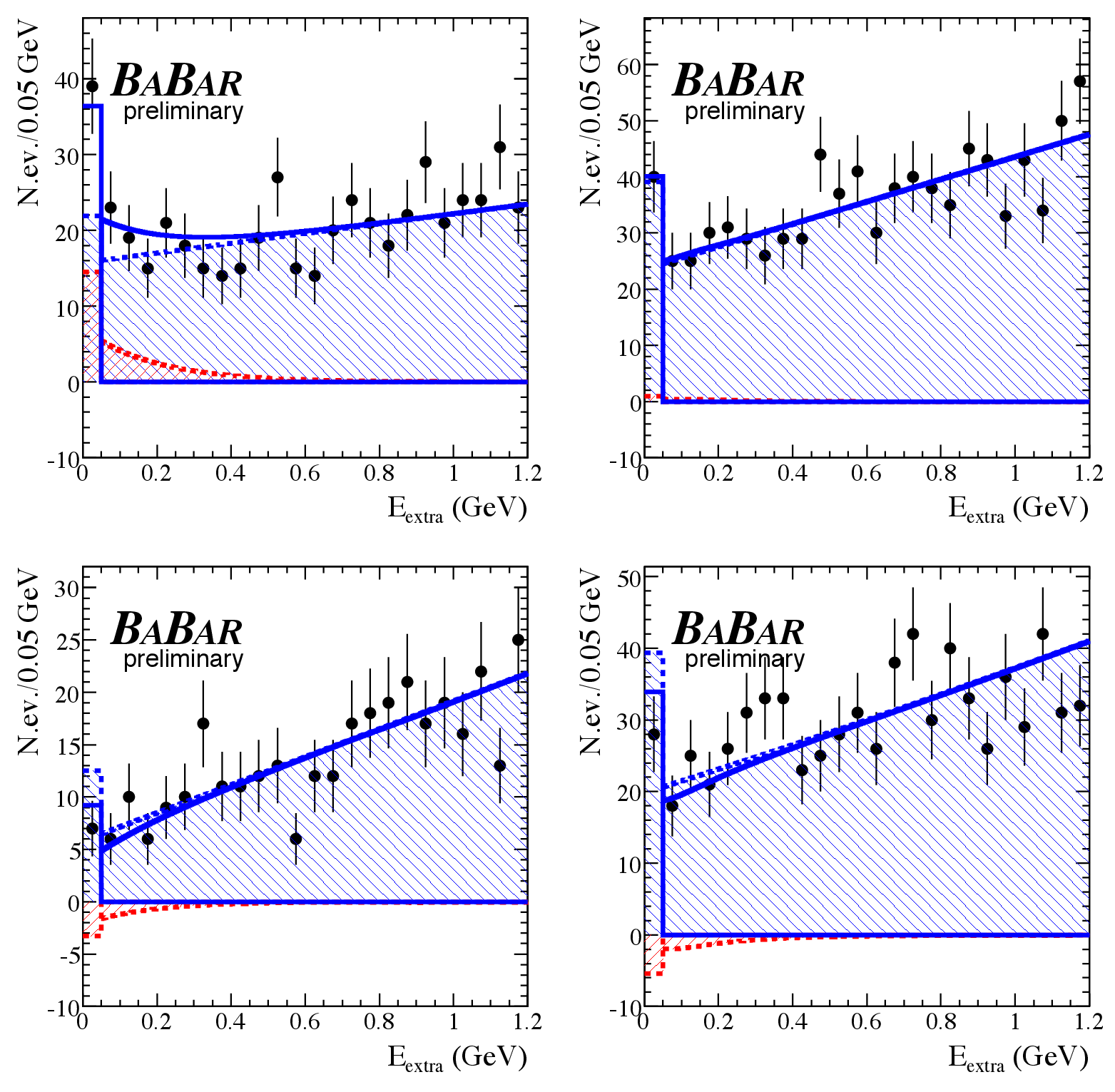}
\caption{Measured $E_{extra}$ spectra in $B \ra K^{*} \nu \bar \nu$ modes: $K^{*0} \ra K^+ \pi^-$ (top left), $K^{*+} \ra K^0_S(\pi^+ \pi^-)  \pi^+$ (top right),  $K^{*+} \ra K^0_S(\pi^0 \pi^0)  \pi^+$ (bottom left), and $K^{*+} \ra K^+  \pi^0$ (bottom right). Curves show the total fit (solid), signal (red dots) and background (hatched area).}
\label{fig:e-extra}
\end{figure}

\babar has searched for $B \ra K^{*0} \nu \bar \nu$ and $B \ra K^{*+} \nu \bar \nu$ ($B \ra K^+ \nu \bar \nu$)
modes in the recoil of semileptonically  tagged $B \ra D^{(*)} \ell \nu$ events using 454~(351)~million $B \bar B$ events. The semileptonic tags are selected by constructing the angle $\theta_{B,D\ell}$ between the $B$ and the $D\ell$ system from

\begin{eqnarray}
\cos \theta_{B,D\ell}&=&\frac{2E_B E_{D\ell}-m^2_B-m^2_{D\ell}}{2|\vec p_B||\vec p_{D\ell}|},
\label{eq-fl-low}
\end{eqnarray}

\noindent
where, $E_B, E_{D\ell}$ are the energies of the $B$ and $D\ell$ system, $m_B, m_{D \ell}$ are corresponding masses and $\vec p_B, \vec p_{D\ell}$ are their momenta. For signal events, the $\cos \theta_{B,D\ell}$ distribution is bounded by the $[-1,+1]$ interval, that may be slightly increased by resolution effects, while a large fraction of background events falls outside this region. The $D^0$ and $D^*$ mesons are reconstructed in several final states. In the recoil we select only $K^+$, $K^+ \pi^-$, $K^0_S \pi^+$, and $K^+ \pi^0$ final states that have  no additional tracks in the event. Due to the two missing neutrinos, we require a large missing energy. In the $K^+ \nu \bar \nu$ analysis signal selection is accomplished with a multivariate analysis \cite{ranfor} that optimizes Punzi's figure of merit \cite{punzi}.

\begin{eqnarray}
PUNZI &= & \frac{N_{sig}}{n_\sigma / 2 + \sqrt{N_{bg}}},
\label{eq-punzi}
\end{eqnarray}

\noindent
where the significance is set to $n_\sigma =3$. In the $K^* \nu \bar \nu$ analysis  Punzi's figure of merit is used to optimize the selection criteria of six variables: $\cos \theta_{B,D\ell}$, the ratio of second-to-zeroth Fox-Wolfram moments $R_2$, the $K^*$ mass $m_{K^*}$, the lepton momentum $p^*_\ell$,  the sum of the missing energy and missing momentum $E^*_{miss}+c p^*_{miss} $, and the polar angle of the missing momentum, where the latter three variables are measured in the CM frame. We fit the extra neutral energy\footnote{energy of all neutral showers in the electromagnetic calorimeter not associated with signal or semileptonic tag.} in the region $\rm 0.05~GeV < E_{extra} < 1.2~GeV$ to the expected signal and background shapes determined from signal and background MC samples, respectively. Figure~\ref{fig:e-extra} shows the observed $E_{extra}$ distributions for the four $K^* \nu \bar \nu$ modes. Semileptonic double tags ($i.e.$ both $B$'s decay to $D^{(*)} \ell \nu$) are used in both analyses as a control sample to validate the simulation.

We see no significant signal yield in any of these modes and set preliminary branching fraction upper limits at $90\%$ confidence level of:

\begin{eqnarray}
{\cal B}(B \ra K^+ \nu \bar \nu) &=& 4.2 \times 10^{-5} ~@ 90\% ~CL, \nonumber \\
{\cal B}(B \ra K^{*0} \nu \bar \nu) &=& 9.0 \times 10^{-5} ~@ 90\% ~CL, \nonumber \\
{\cal B}(B \ra K^{*+} \nu \bar \nu) &=& 21 \times 10^{-5} ~@ 90\% ~CL.
\label{eq-knunu-d}
\end{eqnarray}

\begin{figure}[h]
\centering
\includegraphics[width=80mm]{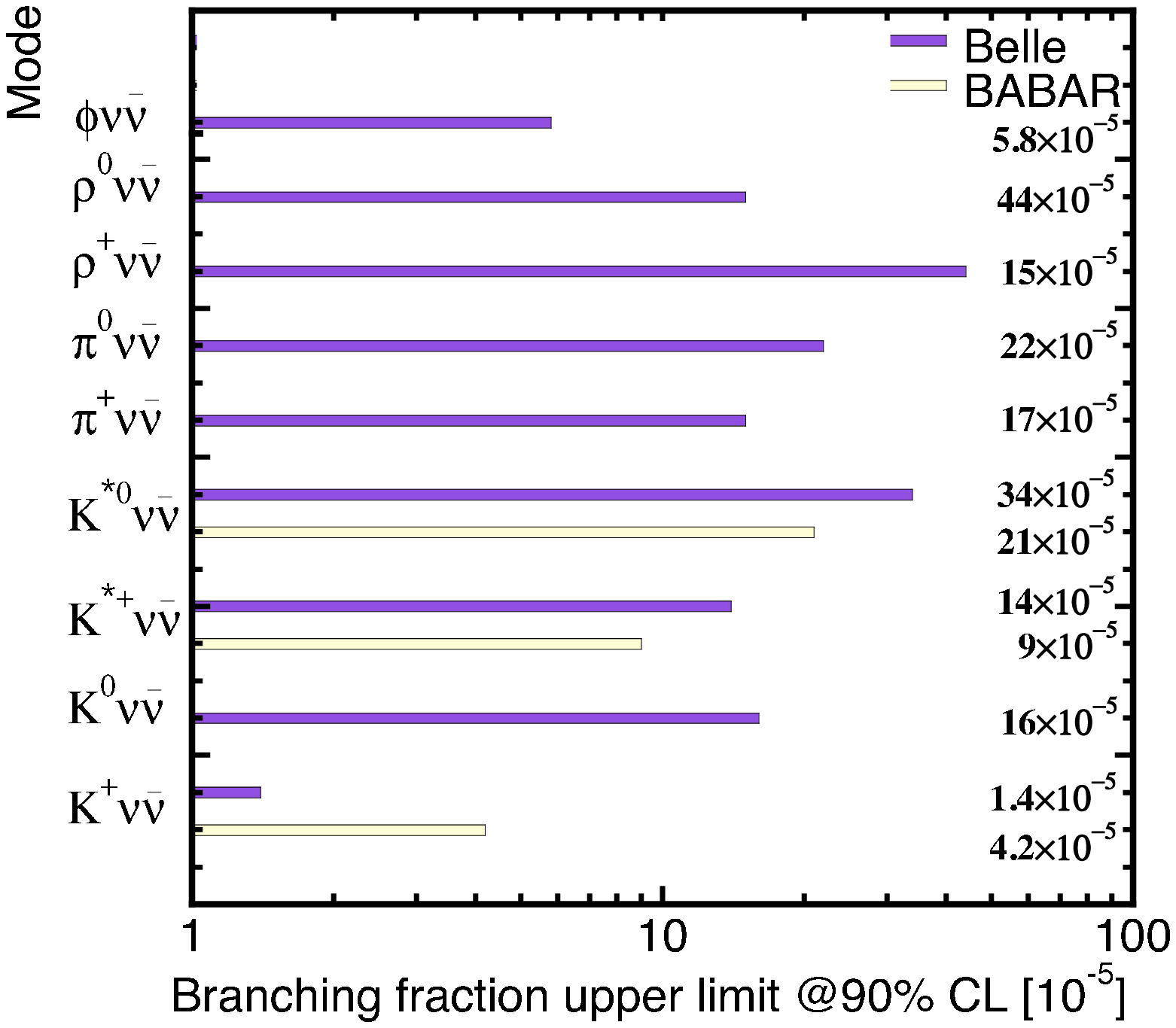}
\caption{Branching fraction upper limits at $90\% $ confidence level for $B \ra h \nu \bar \nu$ modes from \babar (yellow bars) and Belle (dark-hatched bars), where the hadron ($h$) is $K^+, K^{*0}, K^{*+}, \pi^+, \pi^0, \rho^+, \rho^0$, or $\phi$.}
 \label{fig:knunu}
\end{figure}

Figure~\ref{fig:knunu} shows the status of searches for exclusive $(K , K^*, \pi, \rho, \phi) \nu \bar \nu$ modes from \babar and Belle (for 535 million $B \bar B$ events) \cite{belle08a}. The $90\%~CL$ branching fraction upper limits for $K \nu \bar \nu$ ($K^* \nu \bar \nu$) modes are still a factor of four (seven) above the SM prediction. For $B \ra K^* \nu \bar \nu $ the \babar upper limits are the lowest, while for 
$B^+ \ra K^+ \nu \bar \nu$ Belle has set the lowest upper limit of ${\cal B}(B \ra K^+ \nu \bar \nu) < 1.4 \times 10^{-5} @ 90\%~CL$.

\section{Conclusion and Outlook}

The $B^0$ and $B^+$ averaged $B \ra K^{(*)} \ell^+ \ell^-$ total and partial branching fractions from \babar agree well with the SM predictions. In the high $q^2$ and entire $q^2$ regions, isospin asymmetries are consistent with zero as expected in the SM. In the low $q^2$ region, however, we observe large isospin asymmetries in the $K \ell^+ \ell^-$ and $K^* \ell^+ \ell^-$ data. We find no evidence for $K^0_S \ell^+ \ell^-$ events in the low $q^2$ region. While for $K \ell^+ \ell^-$ the fit yields an unphysical central value we measure ${\cal A}_I = -0.56^{+0.17}_{-0.15}\pm 0.03$ for $B \ra K^* \ell^+ \ell^-$. This result is qualitatively more consistent with the flipped-sign $\widetilde C_7$ model than with the SM. Summing the $\Delta \log {\cal L}$ curves for $K \ell^+ \ell^-$ and $K^* \ell^+ \ell^-$ samples we measure an isospin asymmetry of ${\cal A}_I = -0.64^{+0.15}_{-0.14}\pm 0.03$ for the combined sample that excludes the null hypothesis with a significance of $3.9 \sigma$. If isospin asymmetry in the low $q^2$ region expected in the flipped-sign $\widetilde C_7$ model for the combined sample is similar or smaller than that for $K^* \ell^+ \ell^-$, we may need additional sources of isospin violation to explain the \babar measurements. 

All observed \CP asymmetries are consistent with zero. The measured ${\cal R}_{K^{(*)}}$ ratios are consistent with lepton flavor universality. The measurements of the $K^*$ polarization and the lepton forward-backward asymmetry are consistent with the SM prediction. Both \babar and Belle results, however, favor the flipped-sign $\widetilde C_7$ model over the SM. They disfavor the flipped-sign $\widetilde C_9 \widetilde C_{10}$ model at  the $> 3\sigma$ level. To proceed beyond these qualitative statements we need to perform a model-independent global fit to measurements of ${\cal A}_{FB}$, ${\cal F}_L$ and other observables using data from both experiments. In this fit we should include also corresponding measurements of the inclusive $B \ra X_s \ell^+ \ell^-$ channel and results of other electroweak penguin decays, since these might help to improve precision in the extraction of the effective Wilson coefficients. Presently, however, all existing measurements and their SM predictions have too large uncertainties to uncover deviations from the SM. 
Eventually, a global model-independent fit may allow us to extract non-SM contributions (moduli and arguments) in the effective Wilson coefficients $\widetilde C_7$, $\widetilde C_9$ and $\widetilde C_{10}$ if they exist. 

Belle has set the lowest $90\%~CL$ branching fraction upper limit on $B \ra \pi^+ \ell^+ \ell^-$ that lies about a factor of 1.5 above the SM prediction. \babar has set new branching fraction upper limits at $90\%$ CL for  $B \ra K^+ \nu \bar \nu$, $B \ra K^{*0} \nu \bar \nu$ and $B \ra K^{*+} \nu \bar \nu$. For $K^* \nu\bar  \nu$ modes, \babar has set the lowest upper limits, while for $B \ra K^+ \nu \bar \nu$ the Belle upper limit is the lowest. 

The entire \babar data sample consists of 465~ million $B \bar B$ events. Though all present results in the exclusive modes will be updated with the full \babar data sample, the improvements in precision will be rather limited. We may succeed, however, in performing measurements of partial branching fractions, decay-rate asymmetries, longitudinal polarization and lepton forward-backward asymmetry in four rather than two regions of $q^2$. Furthermore, additional information will be obtained from an analysis of the 
sum-of-exclusive modes $B \ra K n\pi \ell^+ \ell^-~(n \leq 3)$ to approximate the inclusive $B \ra X_s \ell^+ \ell^-$ decays, in which we will measure the same observables with the full \babar data set as in the exclusive analyses. A fully-inclusive $B \ra X_s \ell^+ \ell^-$ is presently not feasible, since the kinematic variables $\Delta E$ and $m_{ES}$ are not defined here. In order to reject background from semileptonic decays an alternate strategy is needed. A powerful method consists of reconstructing one $B$ meson completely in a hadronic final state and then look for a lepton pair in the recoil. Since the efficiency for reconstructing hadronic final states is only $\sim 10^{-3}$, a data sample much larger than presently available is needed to utilize this method successfully.

Using the present analysis strategy we estimate about 15 billion $B \bar B$ events to measure $A_{FB}$ in $K^* \ell^+ \ell^-$ in eight bins of $q^2$ with a precision of $\sim 30\%$ in each bin. For a similar analysis of the sum-of-exclusive decays about 7-8 billion $B \bar B$ events should be sufficient, if we reconstruct about half of all final states. A fully inclusive analysis, however, that looks for a dilepton pair from a common vertex in addition to fully reconstructed $B$ in hadronic final states requires at least 50 billion $B \bar B$ events to see of the order of 100 events for an overall efficiency of $5 \times 10^{-4}$. Assuming the SM branching fraction and an overall efficiency of $2.5 \times 10^{-4}$, a sample of 2~billion $B \bar B$ events is needed to observe 10 events in $B \ra K^{*0} \nu \bar \nu$. In similar size $B \bar B$ sample we expect about 10 $\pi^+ \ell^+ \ell^-$ events assuming the SM branching fraction and the performance of the previous \babar analysis. 

The LHCb \cite{lhcb} experiment at CERN will start data taking in 2008/2009 and the KEKB may be gradually upgraded to SuperKEKB \cite{superbelle} in the coming years. LHCb and SuperKEKB will accumulate sufficient data to perform the ${\cal A}_{FB}$ measurements in $B \ra K^* \ell^+ \ell^-$ and in the sum-of-exclusive $B \ra K n \pi \ell^+ \ell^-$ modes. In addition they will observe $B \ra \pi \ell^+ \ell^-$. For the inclusive $B \ra X_s \ell^+ \ell^-$ and $B \ra K^{(*)} \nu \bar \nu$ analyses, however, we need to wait for the high-luminosity SuperB factory \cite{superb} that is proposed at Frascati. With a design luminosity of $10^{36} cm^{-2} s^{-1}$ SuperB will produce more than 10 billion $B \bar B$ events a year.

\begin{acknowledgments}
I would like to thank my \babar collegues D. Doll, K. Flood, P. Jackson, F. Porter and L. Sun for useful discussions. I would also like to thank Th. Feldmann for providing updated branching fraction predictions. This work has been supported by the Norwegian Research Council.

\end{acknowledgments}


\bigskip 

\begin{thebibliography}{99}

\bibitem{buchalla} G.~Buchalla, A.~J.~Buras and M.~E.~Lautenbacher, Rev.\ Mod.\ Phys.\  {\bf 68}, 1125 (1996).
\bibitem{misiak} C. Bobeth, M. Misiak and J. Urban, Nucl. Phys. {\bf B574}, 291 (2000).
\bibitem{greub} H.H Asatryan $et~al.$, Phys. Rev. {\bf D65}, 034009 (2002); Phys. Lett. {\bf B507}, 162, (2001) .
\bibitem{hiller03} G. Hiller and F.Kr\"uger, Phys.Rev. {\bf D69}, 074020 (2004).
\bibitem{beneke01} M. Beneke, Th. Feldmann, and D. Seidel;  Nucl. Phys.{\bf B612}, 25 (2001).
\bibitem{misiak93} M. Misiak, Nucl. Phys. {\bf B393}, 23 (1993); Erratum-ibid. {\bf B439}, 461 (1995).
\bibitem{NewPhysics} G.~Burdman, Phys.\ Rev.\  {\bf D52}, 6400 (1995);
J.~L.~Hewett and J.~D.~Wells, Phys.\ Rev.\  {\bf D55}, 5549 (1997);
W.~J.~Li, Y.~B.~Dai and C.~S.~Huang, Eur.\ Phys.\ J.\  {\bf C40}, 565 (2005);
Y.~G.~Xu, R.~M.~Wang and Y.~D.~Yang,  Phys.\ Rev.\  {\bf D74}, 114019 (2006);
P.~Colangelo, F.~De Fazio, R.~Ferrandes and T.~N.~Pham, Phys.\ Rev.\  {\bf D73}, 115006 (2006);
C.-H.~Chen and C.Q.~Geng, Phys. Rev. D \textbf{66} 094018 (2002).
\bibitem{feldmann02} T.~Feldmann and J.~Matias, JHEP {\bf 0301}, 074 (2003). 
\bibitem{yan} Q.~S.~Yan, C.~S.~Huang, W.~Liao and S.~H.~Zhu, Phys.\ Rev.\  D {\bf 62}, 094023 (2000).
\bibitem{beneke05} M. Beneke, Th. Feldmann, and D. Seidel;  Eur.Phys.J. {\bf C41}, 173 (2005).
\bibitem{ali} A. Ali, P. Ball, L.T. Handoko and G. Hiller, Phys. Rev. {\bf D61}, 074024 (2000);
A.~Ali, E.~Lunghi, C.~Greub and G.~Hiller, Phys.\ Rev.\  {\bf D 66}, 034002 (2002).
\bibitem{ball} P. Ball and Zwicky,  Phys.Rev.{\bf D71}, 014029 (2005). 
\bibitem{defazio} F. De Fazio, Th. Feldmann, T. Hurth, Nucl.Phys. {\bf B733}, 1 (2006), Erratum-ibid. {\bf B800}, 405 (2008); JHEP {\bf 0802|}, 031(2008).
\bibitem{kruger} F. Kruger, L. M. Sehgal, N. Sinha and R. Sinha, Phys. Rev. {\bf D61}, 114028 (2000), [Erratum-ibid. {\bf D63}, 019901 (2001)].
\bibitem{hfag} Heavy Flavor Averaging Group, E. Barberio $et~al.$, arXiv:hep-ex/0704.3575 (2007).
\bibitem{kruger2} F. Kr\"uger and J. Matias, Phys. Rev. {\bf D71}, 094009 (2005).
\bibitem{bobeth08} C. Bobeth, G. Hiller and G. Piranishvili, arXiv:hep-ex/0805.2525 (2008).
\bibitem{stewart} I. W. Stewart and F. J. Tackmann, Phys. Rev. {\bf D75}, 034016 (2007).
\bibitem{huber} T. Huber, T. Hurth, and E. Lunghi,  submitted to Nucl.Phys.B, e-Print: arXiv:0712.3009 [hep-ph].
\bibitem{kim2lu} C.S.~Kim, Y.G.~Kim and  C.D.~Lu, Phys. Rev. \textbf{D64}, 094014 (2001). 
\bibitem{buchalla2} G. Buchalla $et~al.$, Phys. Rev. {\bf D63}, 014015 (2001). 
\bibitem{hou} A. Hovhannisyan, W. S. Hou and N. Mahajan, Phys. Rev. {\bf D 77}, 014016 (2008).
\bibitem{bobeth01} C.~Bobeth, T.~Erweth, F.~Kruger, and J.~Urban, Phys. Rev.  \textbf{D64} 074014 (2001). 
\bibitem{ali2} A. Ali, T.Mannel and T. Morozumi, Phys.Lett. {\bf B273}, 505 (1991). 
\bibitem{BaBarDetector} B.~Aubert {\it et al.}  [\babar Collaboration], Nucl.\ Instrum.\ Meth.\   {\bf A479}, 1 (2002).
\bibitem{PDG} W.~M.~Yao {\it et al.}  [Particle Data Group],  J.\ Phys.\ G {\bf 33}, 1 (2006).
\bibitem{ArgusShape} H.~Albrecht {\it et al.}  [ARGUS Collaboration], Z.\ Phys.\   {\bf C48}, 543 (1990).
\bibitem{feldmann} T.~Feldmann {\it et al.}, private communication (2008).
\bibitem{babar06} B. Aubert {\it et~al.} (\babar collaboration), Phys. Rev. {\bf D73}, 092001 (2006). 
\bibitem{belle2} A. Ishikawa {\it et al.}, (Belle collaboration) Phys.\ Rev.\ {\bf D72}, 092005 (2005). 
\bibitem{cdf} T. Aaltonen  {\it et al.} (CDF collaboration), 	arXiv:0804.3908v1 (2008).
\bibitem{zhong} M.~Zhong, Y.-L.~Wu, W.-Y.~Wang, IJMO {\bf A18}, 1959 (2003).
\bibitem{babar1}  B. Aubert {\it et~al.} (\babar collaboration), Phys. Rev. Lett.{\bf  93}, 081862 (2004). 
\bibitem{belle3} K. Abe \textit{et al.}, (Belle collaboration) BELLE-CONF-0415 [hep-ph/0410006]. 
\bibitem{babar04} B. Aubert {\it et~al.} (\babar collaboration), Phys. Rev. {\bf D70}, 112006 (2004). 
\bibitem{run5afb} B.~Aubert {\it et al.}  (\babar Collaboration),  arXiv:hep-ex/0604007, submitted to PRL.
\bibitem{babar07} B. Aubert {\it et~al.} (\babar collaboration), Phys. Rev. {\bf D76}, 031102 (2007). 
\bibitem{belle} A.~Ishikawa {\it et al.}, (Belle collaboration) Phys.\ Rev.\ Lett.\  {\bf 96}, 251801 (2006). 
\bibitem{ali3} A. Ali and T.Mannel, Phys.Lett. {\bf B264}, 447 (1991);  Erratum-ibid.{\bf B274}, 526 (1992).
\bibitem{belle08} J.-T. Wei, K.-F. Chen \textit{et al.} (Belle collaboration), Phys. Rev.{\bf D78}, 011101 (2008).
\bibitem{babar-pill} B. Aubert \textit{et al.} (\babar Collaboration), Phys. Rev. Lett. 99, 051801 (2007).
\bibitem{ranfor}  I.Narsky arXiv:physics/0507143v1 (2005).
\bibitem{punzi} G. Punzi, Proceedings of Phystat  79 (2003).
\bibitem{belle08a} K.-F. Chen \textit{et al.} (Belle collaboration), PRL {\bf 99}, 221802 (2007).
\bibitem{lhcb} S. Amato \textit{et al.} (LHCb Collaboration), CERN-LHCC-98-04, CERN-LHCC-P-4 (1998); T. Nakada (LHCb collaboration), Acta Phys.Polon.{\bf B38}, 299 (2007). 
\bibitem{superbelle} M. Yamauchi, Nucl. Phys. Proc. Suppl. {\bf 111}, 96 (2002).
\bibitem{superb} M. Bona et al. SLAC-R-856, INFN-AE-07-02, LAL-07-15, e-Print: arXiv:0709.0451 [hep-ex]
(2007).





\end{thebibliography}
\end{document}